\documentclass[12pt,preprint]{aastex}
                                                                                
\newcommand{\eps}[1]{\mbox{log~$\epsilon$(#1)}}
\newcommand\iso[2]{$^{\rm #1}$#2}

\def\cs22964{\mbox{CS~22964-161}}
\def\deg{{$^{\circ}$}}
\def\eg{\mbox{\it e.g.}}
\def\etal{\mbox{et al.}}
\def\ie{\mbox{i.e.}}
\def\kmsec{\mbox{km~s$^{\rm -1}$}}
\def\logg{\mbox{log~{\it g}}}
\def\Msun{\mbox{$M_{\odot}$}}
\def\teff{\mbox{$T_{\rm eff}$}}
\def\vturb{\mbox{$v_{\rm t}$}}
\def\rpro{\mbox{$r$-process}}
\def\spro{\mbox{$s$-process}}
\def\ncap{\mbox{$n$-capture}}
                                                                                
\shorttitle{A Double-Lined Binary CEMP Star}
\shortauthors{Thompson et al.}
                                                                                
\begin{document}
                                                                                
\title{
CS22964-161: A Double-Lined Carbon- and s-Process-Enhanced 
Metal-Poor Binary Star\footnote{
This paper includes data gathered with the 6.5m Magellan and 2.5m 
du Pont Telescopes located at Las Campanas Observatory, Chile.}}
                                                                                
\author{
Ian B. Thompson\altaffilmark{1},
Inese I. Ivans\altaffilmark{1,2}, 
Sara Bisterzo\altaffilmark{3},
Christopher Sneden\altaffilmark{4,1}, \\
Roberto Gallino\altaffilmark{3,5},
Sylvie Vauclair\altaffilmark{6},
Gregory S. Burley\altaffilmark{1}, \\
Stephen A. Shectman\altaffilmark{1},
George W. Preston\altaffilmark{1}
}
                                                                                
\altaffiltext{1}{The Observatories of the Carnegie Institution of Washington,
813 Santa Barbara St., Pasadena, CA 91101; ian,iii,burley,gwp@ociw.edu}
                                                                                
\altaffiltext{2}{Princeton University Observatory, Peyton Hall, Princeton,
NJ 08544}

\altaffiltext{3}{Dipartimento di Fisica Generale, Universita' di Torino,
10125 Torino, Italy; bisterzo,gallino@ph.unito.it}

\altaffiltext{4}{Department of Astronomy and McDonald Observatory, 
The University of Texas, Austin, TX 78712; chris@verdi.as.utexas.edu}

\altaffiltext{5}{Centre for Stellar and Planetary Astrophysics, Monash 
University, PO BOX 28M, Clayton, VIC 3800 Australia}

\altaffiltext{6}{Laboratoire d'Astrophysique de Toulouse et Tarbes 
- UMR 5572 - Universit\'e Paul Sabatier Toulouse III - CNRS, 
14 Av. E. Belin, 31400 Toulouse, France; svcr@ast.obs-mip.fr}

\begin{abstract}
A detailed high-resolution spectroscopic analysis is presented for the 
carbon-rich low metallicity Galactic halo object \cs22964.
We have discovered that \cs22964\ is a double-lined spectroscopic
binary, and have derived accurate orbital components for the system.
From a model atmosphere analysis we show that both components are near 
the metal-poor main-sequence turnoff. 
Both stars are very enriched in carbon and in neutron-capture elements
that can be created in the \spro, including lead.
The primary star also possesses an abundance of lithium close to the 
value of the ``Spite-Plateau''.
The simplest interpretation is that the binary members seen today
were the recipients of these anomalous abundances from
a third star that was losing mass as part of its AGB evolution.
We compare the observed \cs22964\ abundance set with nucleosynthesis
predictions of AGB stars, and discuss issues of envelope stability
in the observed stars under mass transfer conditions, and consider
the dynamical stability of the alleged original triple star.
Finally, we consider the circumstances that permit survival of lithium, 
whatever its origin, in the spectrum of this extraordinary system.
\end{abstract}


%%%%%%%%%%%%%%%%%%%%%%%%%%%%%%%%%%%
\section{INTRODUCTION\label{intro}}
%%%%%%%%%%%%%%%%%%%%%%%%%%%%%%%%%%%

The chemical memory of the Galaxy's initial elemental production in 
short-lived early-generation stars survives in  the present-day 
low-mass, low metallicity halo stars showing starkly contrasting 
abundance distributions.
Metal-poor stars have been found with order-of-magnitude differences in 
lithium contents, large ranges in $\alpha$-element abundances
(from [Mg,Si,Ca,Ti/Fe]\footnote{
We adopt the standard spectroscopic notation (Helfer, Wallerstein, 
\& Greenstein 1959\nocite{hel59}) that for elements A and B, 
\eps{A} $\equiv$ {\rm log}$_{\rm 10}$(N$_{\rm A}$/N$_{\rm H}$) + 12.0, and
[A/B] $\equiv$
{\rm log}$_{\rm 10}$(N$_{\rm A}$/N$_{\rm B}$)$_{\star}$ $-$
{\rm log}$_{\rm 10}$(N$_{\rm A}$/N$_{\rm B}$)$_{\odot}$.
Also, metallicity is defined as the stellar [Fe/H] value.} 
~$<$ 0 to $\sim$+1), non-solar Fe-peak ratios,
and huge bulk variations in neutron-capture (\ncap) abundances.

One anomaly can be easily spotted in medium-resolution 
(R~$\equiv$ $\lambda/\Delta\lambda$~$\simeq$~2000)
spectroscopic surveys of metal-poor stars: large star-to-star 
variations in CH G-band strength, leading to similarly large carbon
abundance ranges.
Carbon-enhanced metal-poor stars (hereafter CEMP $\equiv$ [C/Fe]~$\gtrsim$~+1)
are plentiful at metallicities [Fe/H]~$<$~$-$2, with estimates
of their numbers ranging from $\simeq$14\% (Cohen \etal\ 2005\nocite{coh05})
to $\gtrsim$21\% (Lucatello \etal\ 2006\nocite{luc06}).
The large carbon overabundances are mainly (but not always) accompanied
by large \ncap\ overabundances, which are usually detectable only at higher
spectral resolution (R~$\gtrsim$~10,000).
In all but one of the CEMP stars discovered to date, the \ncap\ abundance
pattern has its origin in slow neutron-capture synthesis (the \spro).
The known exception is CS~22892-052 (\eg, Sneden \etal\ 2003a\nocite{sne03a} 
and references therein), which has [C/Fe]~$\sim$~+1 but an \ncap\ 
overabundance distribution clearly consistent with a rapid 
neutron-capture (the \rpro) origin.
Properties of CEMP stars have been summarized recently in a large-sample 
high-resolution survey by Aoki \etal\ (2007)\nocite{aok07}.
They conclude in part that the abundances of carbon and (\ncap)
barium are positively correlated (their Figure~6), pointing to 
a common nucleosynthetic origin of these elements in many CEMP stars.

\cs22964\ was first noted in the ``HK'' objective prism survey of low 
metallicity halo stars (Beers, Preston, \& Shectman 1992\nocite{bps92}, 
hereafter BPS92).
Using as metallicity calibration the strength of the \ion{Ca}{2} K~line,
BPS92 estimated [Fe/H]~$\simeq$~$-$2.62.
They also found \cs22964\ to be one of a small group of stars with
unusually strong CH G-bands (see their Table~8).
Recently, Rossi \etal\ (2005)\nocite{ros05} analyzed a moderate 
resolution spectrum of this star.
From the extant BVJK photometry and distance estimate they suggested
that \cs22964\ is a subgiant: \teff~=~5750~K and \logg~=~3.3.
Two metallicity estimates from the spectrum were in agreement
at [Fe/H]~=~$-$2.5, and three independent approaches to analysis of the
overall CH absorption strength suggested a large carbon abundance,
[C/Fe]~=~+1.1.

We observed \cs22964\ as part of a high resolution survey of candidate 
low-metallicity stars selected from BPS92.
When we discovered that the star shows very strong features of CH
and \ncap\ species \ion{Sr}{2} and \ion{Ba}{2}, similar to many 
binary blue metal-poor (BMP) stars (Sneden, Preston, \& Cowan
2003b\nocite{sne03b}), we added it to a radial velocity monitoring program. 
Visual inspection of the next observation showed two sets of spectral 
lines, and so an intensive monitoring program was initiated on the du Pont 
and Magellan Clay telescopes at Las Campanas Observatory.
In this paper we present our orbital and abundance analyses of \cs22964.
Radial velocity data and the orbital solution are given in \S\ref{radvel},
and broadband photometric information in \S\ref{phot}.
We discuss the raw equivalent width measurements for the combined-light
spectra and the extraction of individual values for the \cs22964\ primary
and secondary stars in \S\ref{ewcalc}.
Determination of stellar atmospheric parameters is presented in 
\S\ref{model}, followed by abundance analyses of the individual stars
in \S\ref{abboth} and of the  \cs22964\ system in \S\ref{abtotal}.
Interpretation of the large lithium, carbon, and \spro\ abundances in
primary and secondary stars is discussed in \S\ref{history}.
Finally, we speculate on the nature of former asymptotic giant branch 
(AGB) star that we suppose was responsible for creation of the unique
abundance mix of the \cs22964\ binary in \S\ref{agb}.

%%%%%%%%%%%%%%%%%%%%%%%%%%%%%%%%%%%
\section{RADIAL VELOCITY OBSERVATIONS\label{radvel}}
%%%%%%%%%%%%%%%%%%%%%%%%%%%%%%%%%%%

We obtained spectroscopic observations of \cs22964\ with the Clay 6.5-m 
MIKE (Bernstein \etal\ 2003\nocite{ber03}) and du~Pont 2.5-m echelle 
(R~$\simeq$~25,000) spectrographs.
Properties of the two spectrographs are presented at the LCO
website\footnote{http://www.lco.cl}.
The Clay MIKE data have $R$~$\equiv$~$\lambda/\delta\lambda$ and continuous
spectral coverage for 3500~\AA~$\lesssim$~$\lambda$ $\lesssim$~7200~\AA.
The du~Pont data have R~$\simeq$~25,000, and out of the large spectral
coverage of those data we used the region
4300~\AA~$\lesssim$~$\lambda$ $\lesssim$~4600~\AA for velocity measurements.
Exposure times ranged from 1245 to 3500 seconds on the Clay telescope and
3000 to 4165 seconds on the du Pont telescope.
The observations generally consist of two exposures flanked by observations 
of a thorium-argon hollow-cathode lamp.

The Magellan observations were reduced with pipeline software written
by Dan Kelson following the approach of Kelson (2003, 2006).\nocite{kel03} 
\nocite{kel06} Post-extraction processing of the spectra
was done within the IRAF ECHELLE package.\footnote{IRAF is distributed by 
the National Optical Astronomy Observatories, which are operated by the 
Association of Universities for Research in Astronomy, Inc., under 
cooperative agreement with the NSF.} 
The du Pont observations were reduced completely with IRAF ECHELLE software. 

Velocities were initially measured with the IRAF FXCOR package using 
MIKE observations of HD~193901 as a template, and a preliminary 
orbit was derived. 
The three MIKE observations of \cs22964\ obtained at zero velocity 
crossing (phase~$\simeq$~0.11) were then averaged together to define a 
new template, hereafter called the syzygy spectrum. 
This spectrum has a total integration time of 10215~sec and 
S/N~$>$~160 at 4260~\AA.
We remeasured the velocities using this new template with the TODCOR 
algorithm (Zucker \& Mazeh 1994).\nocite{zuc94} 
The cross-correlations covered the wavelength interval 
4130~\AA~$<$~$\lambda$~$<$~4300~\AA.
The syzygy spectrum was also used extensively in our abundance 
analysis (\S\ref{abboth}).

\begin{figure}
\epsscale{1.0}
\plotone{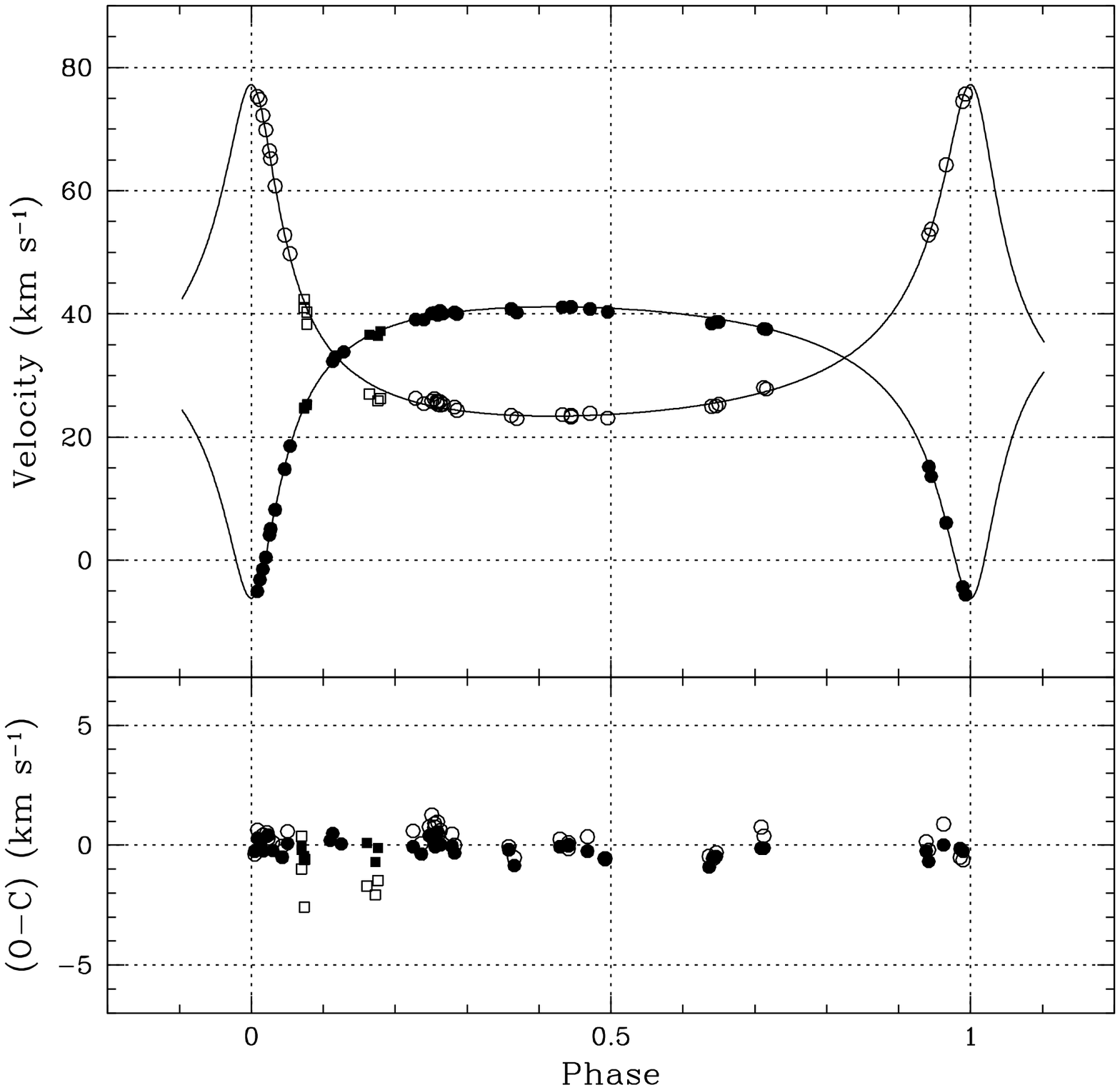}
\caption{
Top panel:  radial velocity observations of \cs22964 primary and 
secondary stars, and best-fit orbital solutions from these velocities.
Circles represent data taken with MIKE on the Magellan Clay telescope 
and squares represent data taken with the echelle spectrograph on the 
du~Pont telescope. 
Filled symbols represent data for the primary and open symbols are for 
the secondary. 
The lines are fits to the Magellan data only.
Bottom panel: differences of the observed velocities and the orbital
solutions.
\label{f1}}
\end{figure}

The radial velocity data were fit with a non-linear least squares solution.
We chose to fit to only the higher resolution and higher S/N Magellan data, 
using the du Pont data to confirm the orbital solution. 
The observations are presented in Table~\ref{tab1} which lists the 
heliocentric Julian Date (HJD) of mid-exposure, the velocities of the primary 
and secondary components, and the orbital phases of the observations. 
The adopted orbital elements are listed in Table~\ref{tab2} and 
the adopted orbit is plotted in Figure~\ref{f1}.
Of particular importance to later discussion in this paper are the
derived masses:  $M_{p}~sin^{3}~i$~= 0.773~$\pm~$0.009~$M_{\odot}$
and $M_{s}~sin^{3}~i~$~= 0.680~$\pm$~0.007~$M_{\odot}$; the orbital
inclination cannot be derived from our data.

We will discuss the velocity residuals to the orbital solution
further in \S\ref{agb}.
Hereafter, references to individual spectra will be by the HJD of
Table~\ref{tab1}, \eg, the spectrum obtained with Magellan MIKE on HJD 
13817.90625 will be called ``observation 13817''.

%%%%%%%%%%%%%%%%%%%%%%%%%%%%%%%%%%%
\section{PHOTOMETRIC OBSERVATIONS\label{phot}}
%%%%%%%%%%%%%%%%%%%%%%%%%%%%%%%%%%%

Preston, Shectman, \& Beers (1991)\nocite{pre91} list $V$~=~14.41, 
$B-V$~=~0.488, and $U-B$~=~$-$0.171 for \cs22964\ based on a single 
photoelectric observation on the du Pont telescope. 
New CCD observations of this star were obtained with the du Pont 
telescope on UT 28 May 2006.
The data were calibrated with observations of the Landolt 
standard field Markarian A (Landolt 1992\nocite{lan92}). 
Observations of the standards and program object were taken at an airmass of 
$\sim$1.12, and we used standard extinction coefficients in the reductions. 
We derived $V$~=~14.43 and $B-V$~=~0.498 for \cs22964.
Old and new photometric data are consistent with the typical $\sim$1\%
observational uncertainties in magnitudes, so we adopt final values
of $V$~=~14.42 and $B-V$~=~0.49.

The ephemeris given in Table~\ref{tab2} predicts a primary eclipse at
HJD2454315.045 (2 August 2007, UT~13.08 hours) and a secondary eclipse at
HJD 2454344.453 (31 August 2007, UT~22.87 hours). \cs22964\ was
monitored on the nights of 2/3 August 2007 and 31 August/1 September
using the CCD camera on the Swope telescope at Las Campanas. Observations
were obtained approximately every hour with a $V$ filter. No
variations in the $V$ magnitude of \cs22964\ in excess of 0.02 
magnitudes were detected.

%%%%%%%%%%%%%%%%%%%%%%%%%%%%%%%%%%%
\section{EQUIVALENT WIDTH DETERMINATIONS\label{ewcalc}}
%%%%%%%%%%%%%%%%%%%%%%%%%%%%%%%%%%%

A double-line binary spectrum is a complex, time-variable mix of two 
sets of absorption lines and two usually unequal continuum flux levels.
Following Preston (1994)\nocite{pre94}\footnote{
A similar kind of analysis of the very metal-poor binary CS~22876-032
has been discussed by Norris, Beers, \& Ryan (2000\nocite{nor00}).}
we define {\it observed} equivalent 
widths $EW_o$ to be those measured in the combined-light spectra, 
and {\it true} equivalent widths $EW_t$ to be those of each star in the 
absence of its companion's contribution.
If the primary-secondary velocity separation is large, one can attempt
to measure observed $EW$s of primary and secondary stars independently.
If this can be accomplished then true $EW$s can be computed from
knowledge of the relative flux levels of the stars.
In practice the derivation of true $EW$s is complex and subject to large
uncertainties.

\begin{figure}
\epsscale{0.9}
\plotone{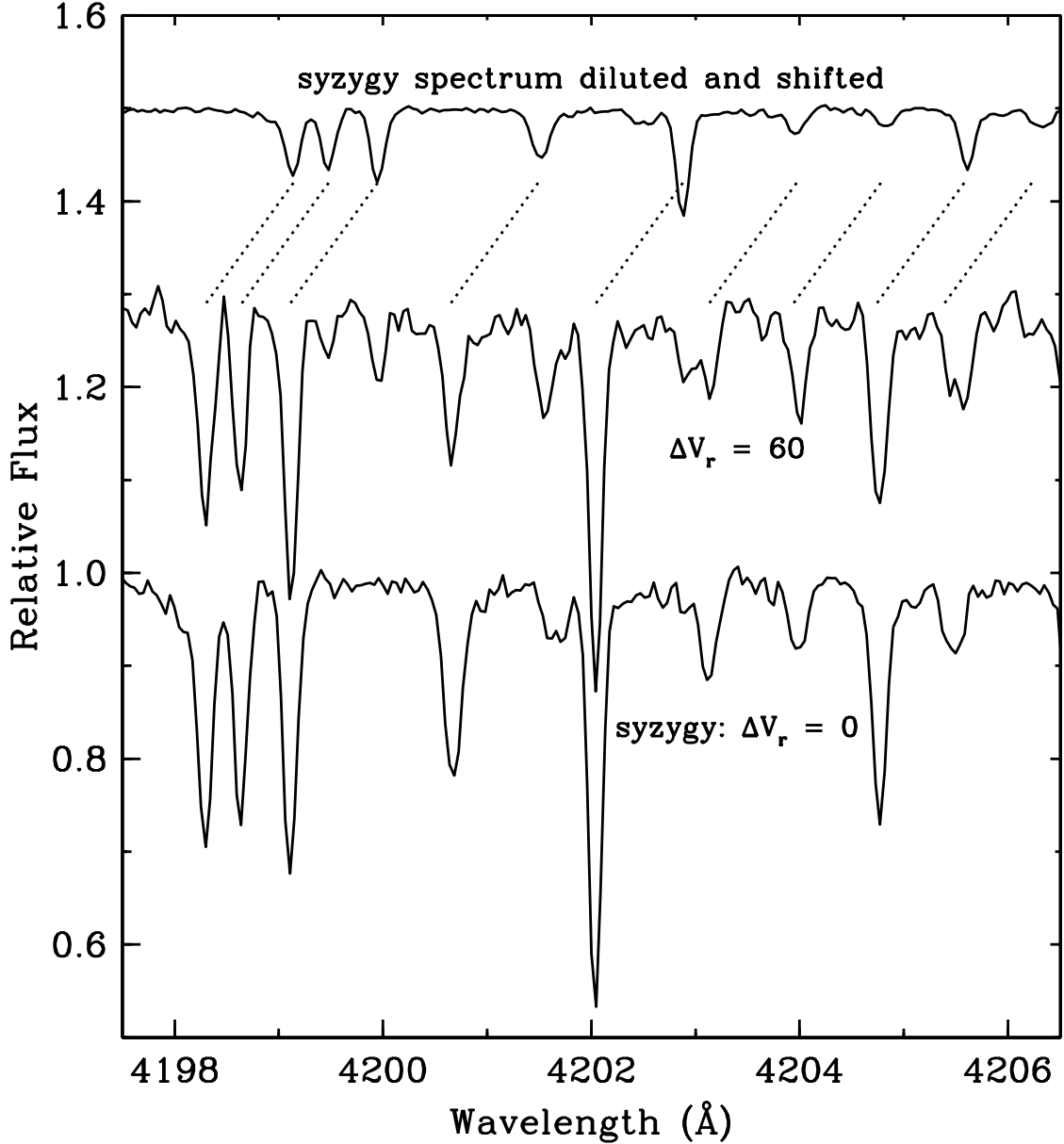}
\caption{
Observed composite absorption spectra of primary and secondary stars.  
The bottom spectrum is the spectrum at syzygy ($\Delta V_R$~= 0~\kmsec); 
this is the mean of three individual observations.
The middle spectrum is the mean of three observations obtained
at phases with $\Delta V_R$~= 60~\kmsec.
The top spectrum represents the original syzygy spectrum after application
of a velocity shift of 60~\kmsec\ and dilution by addition of a continuum 
flux that yields secondary line depths approximately matching those
seen in the middle spectrum.
Dotted lines point to secondary positions of a few lines that are very
strong in the primary spectrum.
The flux scale of the syzygy spectrum is correct, and that other two spectra
have been vertically shifted in the figure for display purposes.
The rest wavelength scale is that of the primary star.
\label{f2}}
\end{figure}

In Figure \ref{f2} we illustrate the difficulties in deducing observed
$EW$s for each star from the observed \cs22964\ composite spectra.
Recall that in the syzygy spectrum (the co-addition of three individual
observations), there is no radial velocity separation, or 
$\Delta V_R$~$\simeq$~0~\kmsec. 
The primary and secondary spectral lines coincide in wavelength, producing 
the simple spectrum shown at the bottom of Figure~\ref{f2}.
In contrast, the middle spectrum in this figure was generated from spectra 
in which the velocity separation was large enough for primary and secondary 
lines to be resolved (hereafter called velocity-split spectra).
We co-added three of these spectra that have similarly large velocity
differences, $\Delta V_R$~$\simeq$~60~\kmsec\ (observations 13580, 
60.12~\kmsec; 13817, 58.09~\kmsec; and 13832, 62.40~\kmsec; 
Table~\ref{tab1}).
The weaker, redshifted secondary absorption lines are obvious from 
comparison of this and the syzygy spectrum.

Identification of the secondary absorption lines is clarified by 
inspection of the top spectrum in Figure \ref{f2}. 
We created this artificial spectrum to mimic the appearance of the
secondary by diluting the observed syzygy spectrum with addition of a 
constant three times larger than the observed continuum, and shifting 
that spectrum redward by 60~\kmsec.
In this line-rich wavelength domain, the velocity shift yields a few 
cleanly separated primary and secondary absorption features. 
For example, 4199.11~\AA\ \ion{Fe}{1} in the primary becomes 
4199.95~\AA\ in the secondary, and observed $EW$s of both stars
can be measured.
More often however, secondary lines are shifted to wavelengths 
very close to other primary lines, destroying the utility of both
primary and secondary features.
An example of this is the \ion{Fe}{1} 4198.33~\AA\ line, which at 
$\Delta V_R$~$\simeq$~60~\kmsec\ becomes 4199.17~\AA\ for the secondary
star, which contaminates the \ion{Fe}{1} 4199.11~\AA\ line of the 
primary star. 
The various blending issues, combined with the intrinsic weakness of 
the secondary spectrum, yield few spectral features with clean
$EW$ values for both primary and secondary stars.

\subsection{Equivalent Widths from Comparison of Syzygy and Velocity-Split
            Spectra\label{ewsyzygy}}

In view of the blending issues outlined above, we derived the 
observed $EW$s through a spectrum difference technique. 
On the mean syzygy spectrum, we measured $EW_{o,tot} = EW_{o,p} + EW_{o,s}$,
where subscripts $o$ denotes an observed $EW$, and $p$ and $s$ denote
primary and secondary stars.  
On five spectra with large primary-secondary velocity separations (13580,
13585, 13587, 13817, 13834) we measured the primary star lines, $EW_{o,p}$.
These five independent values were then averaged, and then the secondary
star's values were computed as $EW_{o,s} = EW_{o,tot} - <EW_{o,p}>$.
In this procedure we attempted to avoid primary lines that would have 
significant contamination by secondary lines in the velocity-split spectra.

The observed $EW$ values are given in Table~\ref{tab3}, along with the line
excitation potentials and transition probabilities.
In this table we also include the atomic data for lines from which 
abundances ultimately derived from synthetic spectrum rather than 
$EW$ computations.  
For one estimate of the uncertainty in our $EW$ measurement procedure, 
we computed standard deviations for the primary-star measurements 
of each line $\sigma EW_{o,p}$.
The mean and median of these values for the whole line data set were
5.2 and 4.1~m\AA, respectively.
The uncertainties in $EW_{o,tot}$ are expected to be smaller because the
syzygy spectrum is the mean of three individual observations and is thus
of higher S/N.
Therefore we take 4$-$5~m\AA\ as an estimate of the uncertainty in $EW_{o,s}$
values determined in the subtraction procedure.

\begin{figure}
\epsscale{0.9}
\plotone{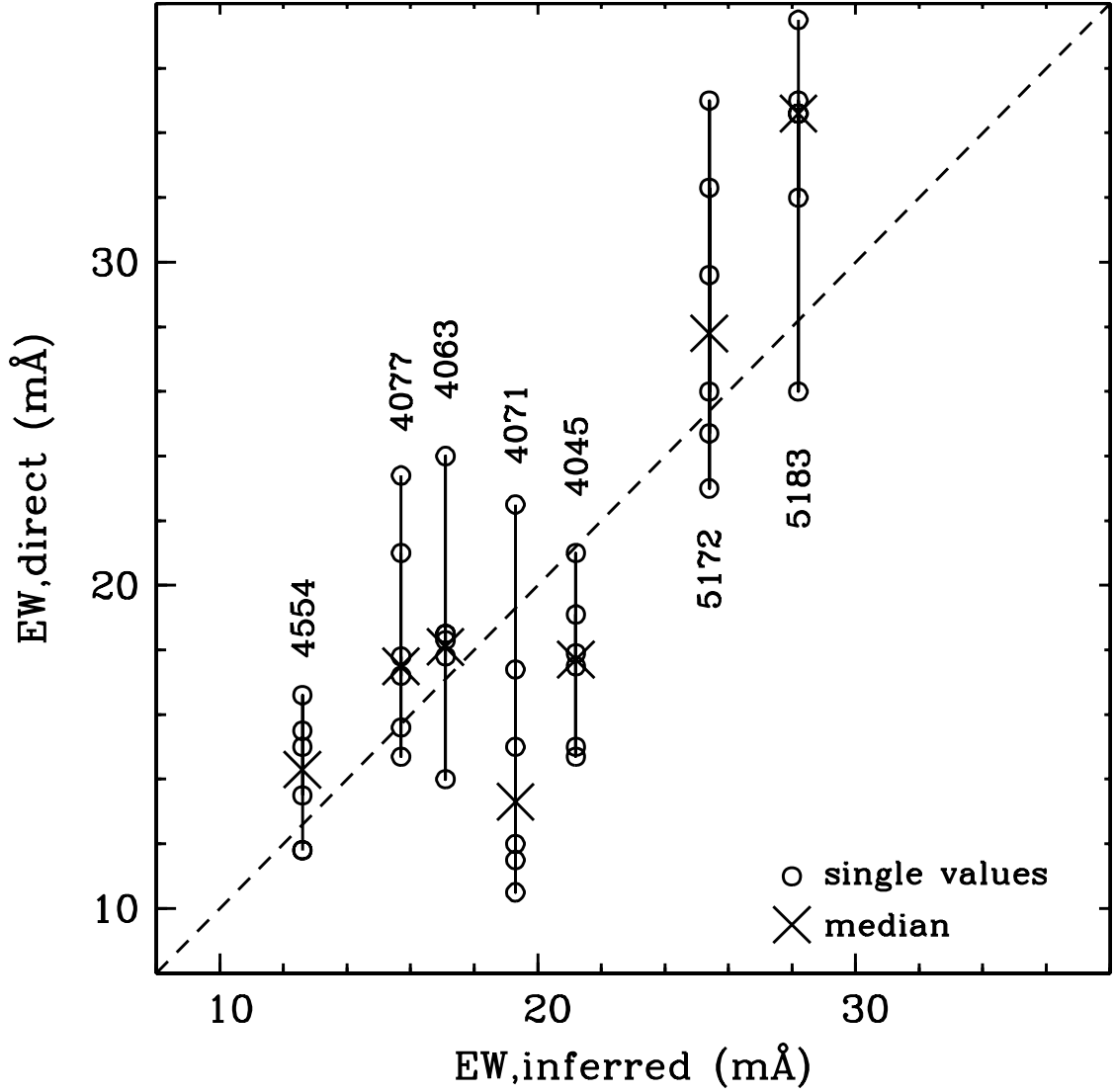}
\caption{
Comparison of observed equivalent widths of the secondary star inferred 
from the syzygy spectrum and those measured directly from the six spectra 
with large velocity splits between primary and secondary stars.
Measurements for each spectral feature are connected by vertical lines 
and labeled with the feature wavelength.
Medians of the $EW$s are displayed as $\times$ symbols.
The dashed slanting line indicates equality of $EW$ values.
\label{f3}}
\end{figure}

For a few very strong transitions the secondary lines are deep enough to
cleanly detect when they split away from the primary lines.
We have employed these to assess the reliability of the $EW$s derived
by the subtraction method described above.
In Figure~\ref{f3} we show a comparison of $EW_{o,s}$ values given
in Table~\ref{tab3} for seven lines that we measured on up to six 
velocity-split spectra (the five named in the previous paragraph 
plus 13832, which was not used in the subtraction procedure).
Taking the means of the $EW_{o,s}$ measurements of each line and comparing
them to the subtraction-based values of Table~\ref{tab3}, the average
difference for the seven lines is 0.8~m\AA\ with a scatter 
$\sigma$~=~3.5~m\AA\ (the median difference is 0.6~m\AA).
This argues that in general the individual $EW_{o,s}$ values agree with the 
$EW_{o,s}$ subtraction-based ones.

Derivation of true $EW$s depends on knowledge of the relative 
luminosities of primary and secondary stars.
Formally, from Preston's (1994)\nocite{pre94} equations 4$-$5, we have
$EW_{p,t} = EW_{p,o}(1 + l_s/l_p)$, and
$EW_{s,t} = EW_{s,o}(1 + l_p/l_s)$, where subscript $t$ represents the
true $EW$, and $l$ denotes an apparent luminosity.
The luminosity ratios will be wavelength-dependent if the two stars are
not identical in temperature.  
For the entire line data set, ignoring weaker primary lines (those
with $EW_{p,o}$~$<$~25.0~m\AA), we calculated a median equivalent 
width ratio ($EW_{p,o}/EW_{s,o}$)~=~5.2.
Inspection of the velocity-split spectra suggested that {\it relative}
line strengths in the secondary spectrum were not radically different 
from those of the primary spectrum.  
Therefore we concluded that the spectral types of the stars are not too 
dissimilar and therefore we concluded that to first approximation
$EW_{p,t}$~$\sim$~$EW_{s,t}$.
This assumption then leads to $EW_{p,o}/EW_{s,o}$~$\sim$ $l_p/l_s$~$\sim$~5.

To account roughly for the small derived temperature difference 
(see \S\ref{model}) we finally adopted $l_p/l_s$~=~5.0 in the 
photometric $V$ bandpass ($\lambda \simeq 5500$~\AA), and increased 
the ratio linearly by a small amount with decreasing wavelength. 
Thus in the $B$ bandpass ($\lambda \simeq 4400$~\AA) we used
$l_p/l_s$~=~5.6.
Final true $EW$ values using this prescription are given in Table~\ref{tab3}.
The correction factors between observed and true $EW$s were approximately 
1.2 and and 6.0 for primary and secondary stars in the $V$ spectral region, 
and the disparity in these factors is larger at $B$.
Clearly the $EW_{p,t}$ values of Table~\ref{tab3} are much more reliably
determined than the $EW_{s,t}$ ones.

%%%%%%%%%%%%%%%%%%%%%%%%%%%%%%%%%%%
\section{STELLAR ATMOSPHERE PARAMETERS\label{model}}
%%%%%%%%%%%%%%%%%%%%%%%%%%%%%%%%%%%

\subsection{Derivation of Parameters\label{params}}

We used the observed $B-V$ color and minimum masses for the binary 
together with the Victoria-Regina stellar models (VandenBerg, 
Bergbusch, \& Dowler 2006)\nocite{van06}, to estimate initial model 
atmospheric parameters for the component stars of \cs22964.
We adopted $B-V$~=~0.49 from \S\ref{phot} and $E(B-V)$~=~0.07 (BPS92; 
a value we also estimate employing the dust maps of Schlegel, Finkbeiner, 
\& Davis 1998)\nocite{sch98} to obtain $(B-V)_{0}$ = 0.42. 
We assumed initially that $sin^{3}~i$~$\simeq$~1.0 for the binary orbit
and thus $M_{p}$~= 0.773~$M_{\odot}$ and $M_{s}$~= 0.680~$M_{\odot}$, 
as derived in \S\ref{radvel}.
We interpolated the Victoria-Regina models computed for [Fe/H]~=~$-$2.31,
[$\alpha$/Fe]~=~+0.3, and $Y$~=~0.24, to obtain evolutionary tracks
for these masses.
These tracks were used to derive $B-V$ colors and luminosity ratios for
the component stars.
We adopted starting values of \teff, \logg, and $l_p/l_s$ where
the tracks give $B-V$~=~0.42 for the combined system.
These values were \teff$_{,p}$~=~6050~K, \teff$_{,s}$~=~5950~K 
and \logg$_p$~=~3.6, \logg$_s$~=~4.2. 
The luminosity ratios using these parameters were $l_p/l_s~(B)$~=~8.65 
and $l_p/l_s~(V)$~=~8.39, somewhat larger than implied by our spectra.

Final model atmospheric parameters were found iteratively from the $EW$ 
data for the two stars.
We employed the LTE line analysis code MOOG\footnote{
Available at: http://verdi.as.utexas.edu/moog.html}
(Sneden 1973)\nocite{sne73} and
interpolated model atmospheres from the Kurucz 
(1998)\nocite{kur98}\footnote{ 
Available at: http://kurucz.harvard.edu/} grid computed with no convective 
overshoot (as recommended by Castelli, Gratton, \& Kurucz 1997\nocite{cas97},
and by Peterson, Dorman, \& Rood 2001\nocite{pet01}).  

\begin{figure}
\epsscale{0.9}
\plotone{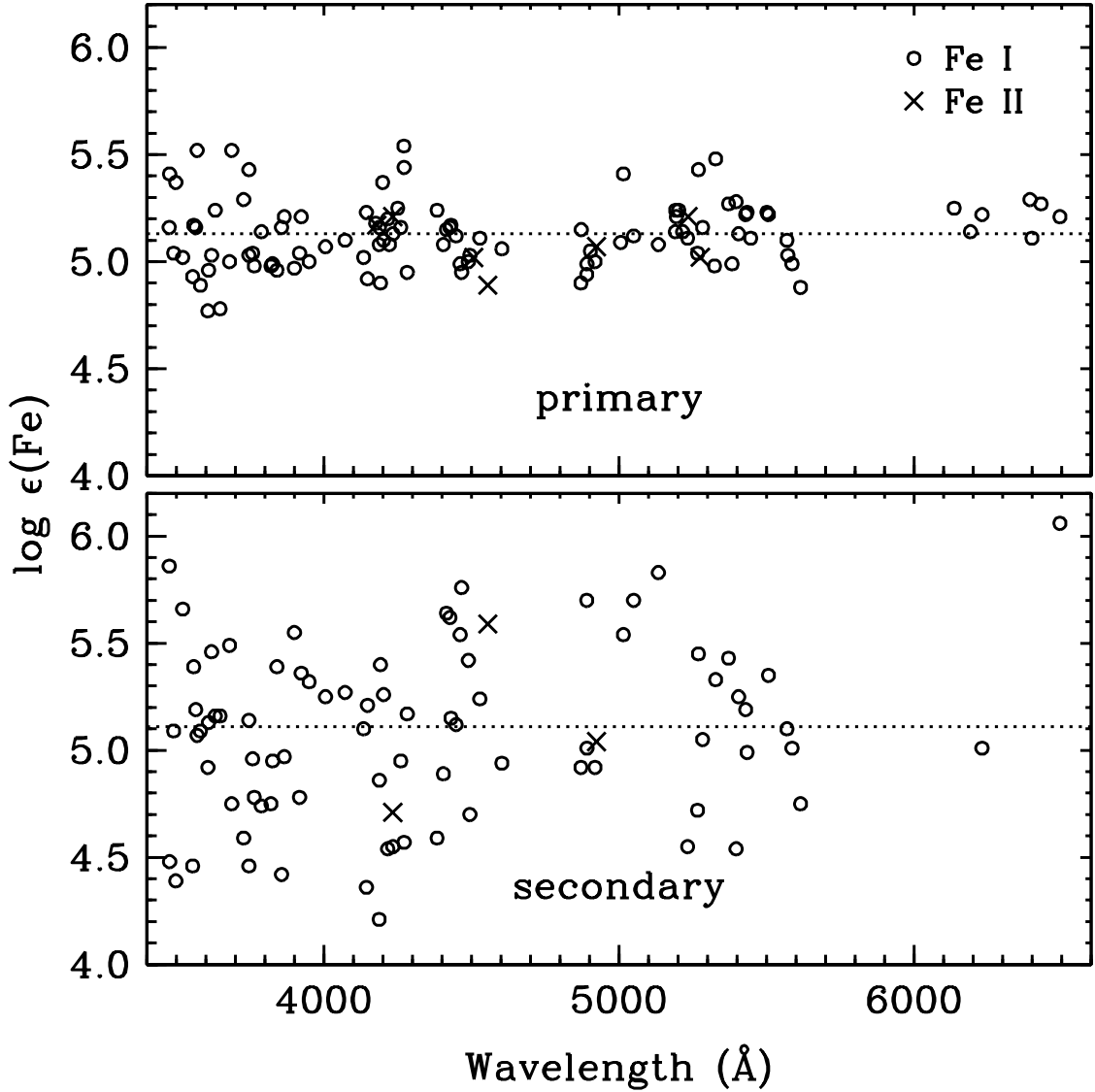}
\caption{
Iron abundances plotted as a function of wavelength for primary (top panel) 
and secondary (bottom panel) stars.  
The symbols are defined in the figure legend.
Dashed horizontal lines are drawn to indicate the mean \ion{Fe}{1}
abundances of primary and secondary stars, \eps{Fe}~=~5.13 and 5.11,
respectively.
\label{f4}}
\end{figure}

We began by using standard criteria to estimate the model parameters of 
the \cs22964\ primary:  {\it (a)} for \teff, no trend of derived \ion{Fe}{1}
individual line abundances with excitation potential; {\it (b)} for
\vturb, no trend of \ion{Fe}{1} abundances with $EW$; {\it (c)} for
\logg, equality of mean \ion{Fe}{1} and \ion{Fe}{2} abundances (for no
other element could we reliably measure lines of the neutral and ionized
species); and {\it (d)} for model metallicity [M/H], a value roughly
compatible with the Fe and $\alpha$ abundances.
These criteria could be assessed reliably for the primary star because
it dominates the combined light of the two stars.
In the top panel of Figure~\ref{f4} we illustrate the line-to-line
scatter and (lack of) trend with wavelength of the primary's \ion{Fe}{1}
and \ion{Fe}{2} transitions.
With iteration among the parameters we derived
(\teff$_{,p}$, \logg$_p$, \vturb$_p$, [M/H]$_p$)~= 
(6050$\pm$100~K, 3.7$\pm$0.2, 1.2$\pm$0.3~\kmsec, $-$2.2$\pm$0.2).

To estimate model parameters for the secondary we assumed that 
derived [Fe/H] metallicities and abundance ratios of the lighter 
elements (Z~$\leq$~30) in the primary and secondary stars should be 
essentially identical if they were formed from the same interstellar cloud.
Iteration among several sets of (\teff$_{,s}$, \logg$_s$) pairs was done
until [Fe/H]$_s$~$\simeq$~[Fe/H]$_p$ and Fe ionization equilibrium
for the secondary was achieved.
Given the weakness of the secondary spectrum and the resulting large 
correction factors used to calculate $EW_{s,t}$ from $EW_{s,o}$ values,
it is not surprising that the spectroscopic constraints on the
secondary parameters were weak.
This is apparent from the adopted Fe abundances displayed in the 
bottom panel of Figure~\ref{f4}.
The $\sigma$ values for individual line abundances were about three times 
larger for the secondary than the primary (Table~\ref{tab4}), 
and number of transitions was substantially smaller (\eg, we measured 
seven \ion{Fe}{2} lines for the primary but only three for the secondary).

We adopted final model parameters of 
(\teff$_{,s}$, \logg$_s$, \vturb$_s$, [M/H]$_s$)~= 
(5850~K, 4.1, 0.9~\kmsec, $-$2.2).
Model uncertainties for the secondary were not easy to estimate and, 
of course, were tied to our opening assumption that the two stars have
identical overall metallicities and abundance ratios.
Thus with [M/H]$_p$~$\equiv$~[M/H]$_s$, the uncertainties in \teff,
\logg, and \vturb\ of the secondary are approximately double their
values quoted above for the primary.  
Thus caution is obviously warranted in interpretation of the model
parameters of the \cs22964\ secondary.

\subsection{Comparison to Evolutionary Tracks\label{evol}}
                                                                                
The well-determined \teff$_{,p}$ and \logg$_p$ values can be used
with the mass and luminosity ratios of the stars to provide an
independent estimate of \teff$_{,s}$ and \logg$_s$.  
Standard relations $L \propto R^2\teff^4$ and $g \propto M/R^2$ lead to
$${\rm log}(L_s/L_p) = {\rm log}(l_s/l_p) = 
{\rm log}(M_s/M_p) + 4\times{\rm log}(\teff_{,s}/\teff_{,p}) - 
{\rm log}(g_s/g_p)$$
Taking an approximate average luminosity ratio to be 
$l_p/l_s$~$\simeq$~5.3, adopting $M_s/M_p$~=~1.15 (Table~\ref{tab2})
and assuming \teff$_{,p}$~=~6050~K and \logg$_p$~=~3.7 from above, 
we get a predicted temperature-gravity relationship 
\logg$_s$~$\simeq$~4$\times$log(\teff$_{,s}$)~$-$~10.8 .

\begin{figure}
\epsscale{0.9}
\plotone{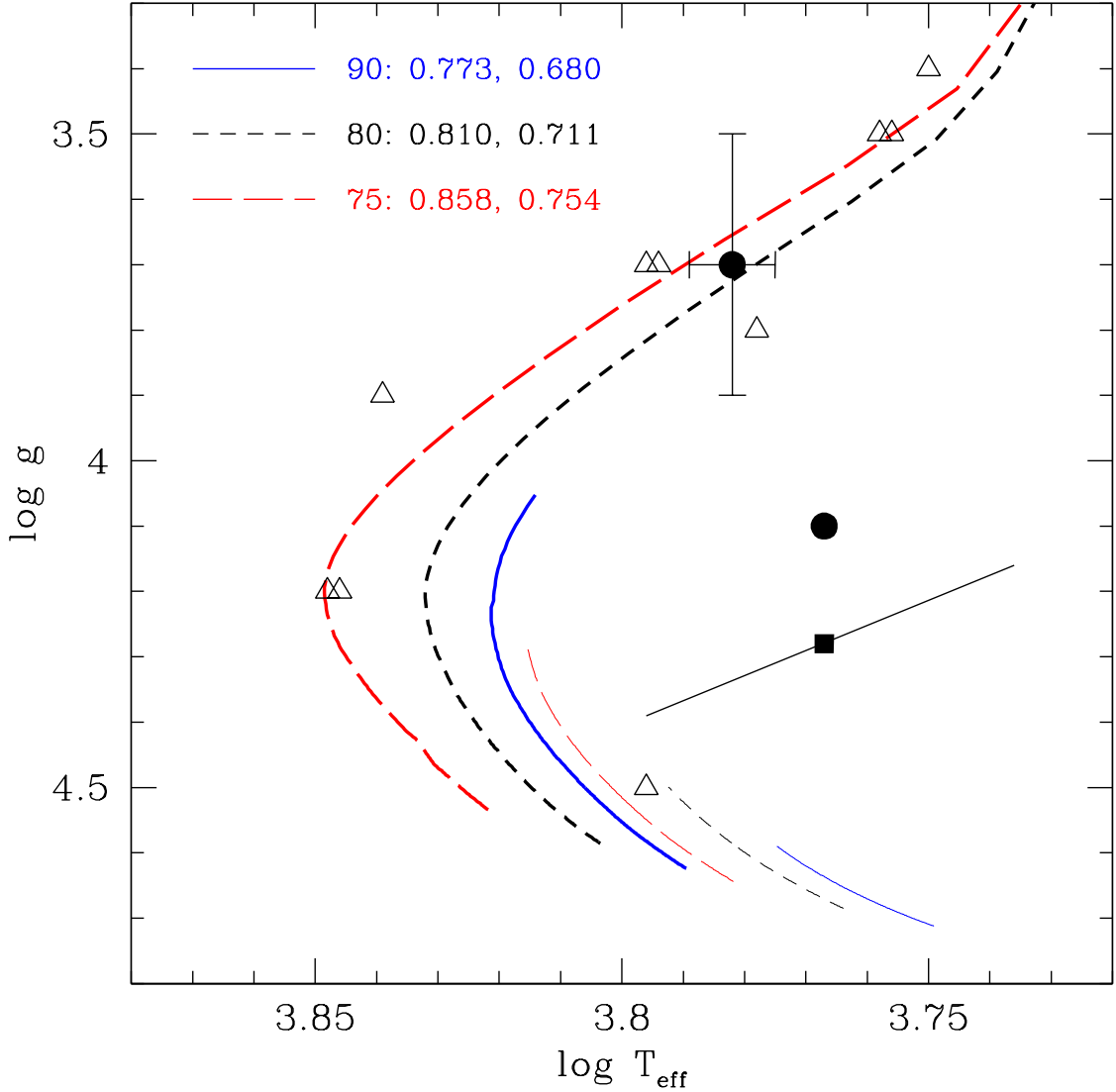}
\caption{
Evolutionary tracks (VandenBerg \etal\ 2006) and CEMP data plotted in the 
log~\teff~$-$~\logg\ plane near the metal-poor main-sequence turnoff.
The chosen tracks correspond to three different inclinations of the binary;
heavy and light lines are for the primary and secondary, respectively.
A full track covers an age range of 4 to 14~Gyr.  
The figure legend identifies the masses of the tracks for three different 
\cs22964\ orbital $sin\ i$ choices, using the notation 
($sin\ i$: $M_{p}$, $M_{s}$).
Filled circles represent the \cs22964\ primary and secondary 
(log~\teff, \logg) values derived from the spectroscopic analysis 
(\S\ref{params}).
The plotted error bars for the primary are also from the spectroscopy.
The square shows the secondary parameters from the dimensional analysis 
described in \S\ref{evol},
The line about this point corresponds to $\Delta$\teff~= $\pm$400~K. 
The triangles represent CEMP stars from the literature; see \S\ref{evol} 
for details.
The triangles for three pairs of stars with identical temperatures and 
gravities have been shifted apart from each other for display purposes. 
\label{f5}}
\end{figure}

In Figure~\ref{f5} we show the very metal-poor main sequence 
turnoff region of the H-R diagram in (log~\teff, \logg) units.
The Victoria-Regina evolutionary tracks (VandenBerg \etal\ 
2006\nocite{van06}; [Fe/H]~=~$-$2.31, [$\alpha$/Fe]~=~+0.3, 
and $Y$~=~0.24) discussed in \S\ref{params} are again employed.
However, the unknown binary orbital inclination of \cs22964\ 
cannot be ignored here.
Therefore we have plotted the tracks for three pairs of masses 
corresponding to assumed $sin\ i$ values of 90\deg, 80\deg, and 75\deg.
Also plotted is a straight line representing the \cs22964\ secondary star
temperature-gravity equation derived above.

The \cs22964\ primary and secondary and (\teff, \logg) positions are
indicated with filled  circles in Figure~\ref{f5}.
We also add data indicated with open triangles from CEMP high 
resolution spectroscopic studies for C-rich stars of similar metallicity, 
taken here to be [Fe/H]~=~$-$2.4~$\pm$~0.4.
The studies include those of Aoki \etal\ (2002)\nocite{aok02},
Sneden \etal\ (2003b)\nocite{sne03b}, and
Cohen \etal\ (2006)\nocite{coh06}.
If \teff$_{,s}$~=~5850~K from the spectroscopic analysis then the 
temperature-gravity relationship from above predicts \logg$_s$~=~4.3 
(indicated by a filled square in the figure). 
Our spectroscopic value of \logg$_s$~=~4.1 lies well within the 
uncertainties of both estimates.

Given the apparently anomalous position of the \cs22964\ secondary
in Figure~\ref{f5}, it is worth repeating its abundance analysis 
using atmospheric parameters forced to approximately conform with the
evolutionary tracks.
This is equivalent to attempting a model near the high temperature,
high gravity end of its predicted relationship shown in the figure.
Therefore we computed abundances for a model with parameter set
(\teff$_{,s}$, \logg$_s$, [M/H]$_s$)~= (6300~K, 4.5, $-$2.2).
Assumption of a secondary microturbulent velocity
\vturb$_s$~= \vturb$_p$~=~ $-$1.2~\kmsec, yields \eps{Fe}~=~5.45 
and 5.27 from \ion{Fe}{1} and \ion{Fe}{2} lines, respectively.
These values are substantially larger than the mean abundances for
primary and secondary given in Table~\ref{tab4}, \eps{Fe}~=~5.11.
This is in agreement with expectations of a larger derived 
metallicity from the 350~K increase in \teff$_{,s}$ for this test.
However, the \ion{Fe}{1} line abundances for this hotter model also
exhibit an obvious trend with $EW$.
This problem could be corrected by increasing \vturb\ to 2.4~\kmsec, 
and then we get \eps{Fe}~=~5.18 and 5.15 from \ion{Fe}{1} and 
\ion{Fe}{2} lines, very close to our final adopted Fe abundances for 
the secondary.
However, it is difficult to reconcile such a large \vturb\ with
the much smaller value determined with more confidence in the primary,
as well as standard values determined in many literature studies of 
near-turnoff stars.

%%%%%%%%%%%%%%%%%%%%%%%%%%%%%%%%%%%
\section{ABUNDANCES OF THE INDIVIDUAL STARS\label{abboth}}
%%%%%%%%%%%%%%%%%%%%%%%%%%%%%%%%%%%

With the $EW$ data of Table~\ref{tab3} and the interpolated model 
atmospheres described in \S\ref{model}, we determined abundances of 
a few key elements whose absorption lines are detectable in both 
primary and secondary observed spectra.
These abundances are given in Table~\ref{tab4}.
They suggest that in general the abundance ratios of all elements in
the two stars agree to within the stated uncertainties. 
Both stars are relatively enriched in the $\alpha$ elements:
$<$[Mg,Ca,Ti/Fe]$_p>$~$\simeq$ +0.4 and $<$[Mg,Ca,Ti/Fe]$_s>$~$\simeq$ +0.3.
Both stars have solar-system Ni/Fe ratios, probably no substantial
depletions or enhancements of Na (special comment on this element
will be given in \S\ref{abtotal}), and large deficiencies of Al.
All these abundance ratios are consistent with expectations for 
normal metal-poor Population~II stars.

More importantly, we find very large relative abundances of C, Sr, 
and Ba ([X/Fe]~=~+0.5 to +1.5; Table~\ref{tab4}) in both primary 
and secondary stars of the \cs22964\ system.
For Sr and Ba abundances we first used the $EW$ subtraction technique, which 
suggested roughly equal abundances of these elements in both stars.
We confirmed and strengthened this result through synthetic spectrum 
computations of the strong \ion{Sr}{2} 4077.71, 4215.52~\AA\ and the 
\ion{Ba}{2} 4554.04~\AA\ lines in the six velocity-split spectra.
To produce the binary syntheses we modified the MOOG line analysis code 
to compute individual spectra for primary and secondary stars, then to 
add them after {\it (a)} shifting the secondary spectrum in wavelength 
to account for the velocity difference between the stars, and 
{\it (b)} weighting the primary and secondary spectra by the 
appropriate luminosity ratio.

\begin{figure}
\epsscale{0.9}
\plotone{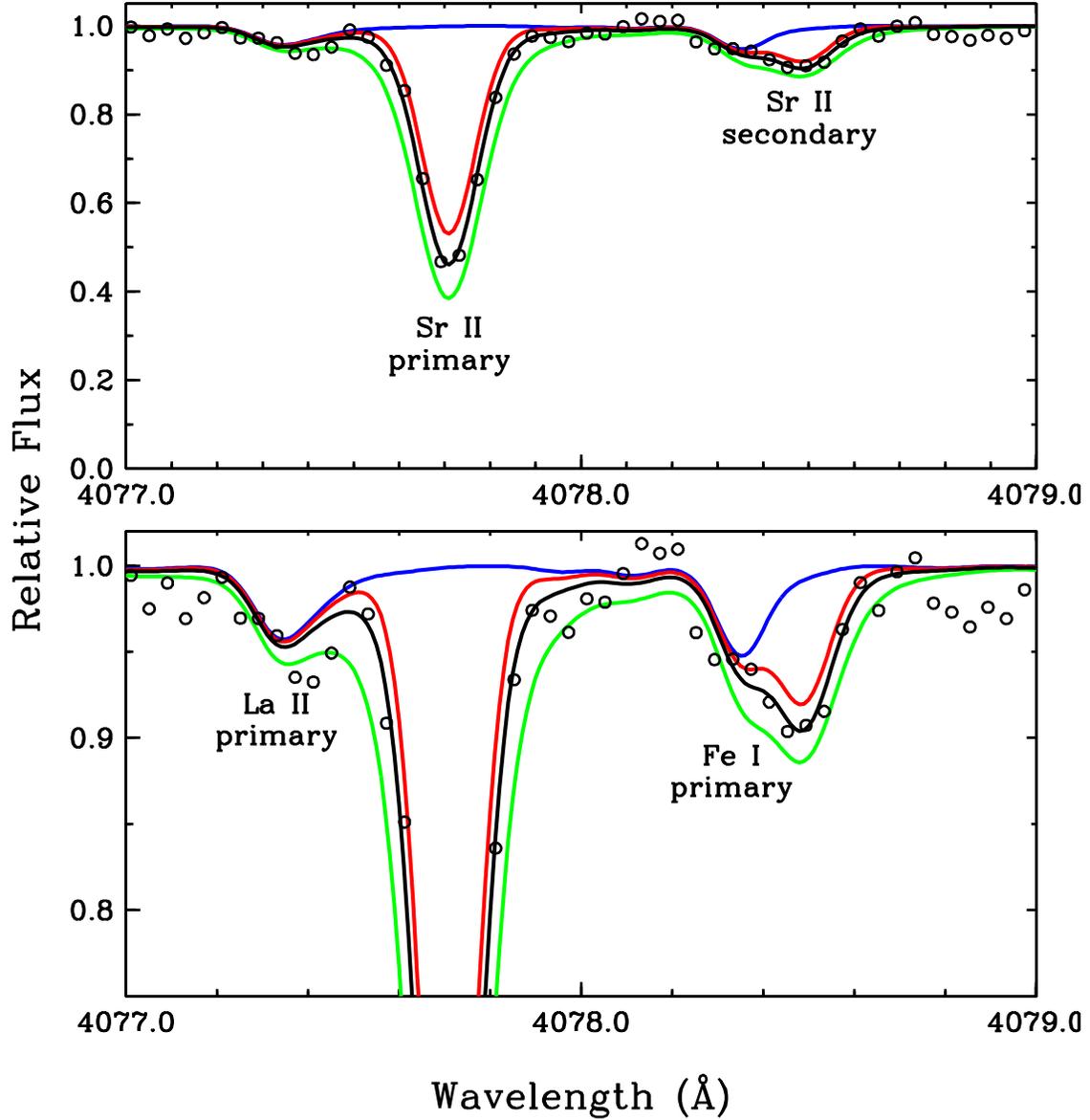}
\caption{
Observed and synthetic spectra of the \ion{Sr}{2} 4077.71~\AA\
line in observation 13817 of \cs22964\ ($\Delta V_R$~$\simeq$~58.09~\kmsec).
The top spectrum shows the full relative flux scale, and the bottom spectrum
covers just the relative flux region above 0.75.
The wavelength scale is at rest velocity for the primary star.
The observed spectrum is depicted with open circles.
The blue line represents a synthetic spectrum computed without any
contribution from Sr.
The black line shows the best overall fit to primary and secondary
\ion{Sr}{2} features: \eps{Sr}~= 0.98 and 0.78, respectively.
The red and green lines indicate syntheses for each star that are 0.5~dex 
smaller and larger, respectively, than the best fits.
\label{f6}}
\end{figure}

In Figure~\ref{f6} we show the resulting observed/synthetic binary 
spectrum match for the \ion{Sr}{2} 4077.71~\AA\ line in observation 13817.
The absorption spectrum of the secondary star is shifted by +58.1~\kmsec\
(+0.79~\AA), in agreement with the observed feature in Figure~\ref{f6}.
The depth of the \ion{Sr}{2} line in the secondary is weak, as expected due
to the luminosity difference between the two stars ($l_p/l_s$~$\simeq$ 6)
at this wavelength.
Note also in Figure~\ref{f6} the relative insensitivity of the
feature to abundance changes for the secondary, even with the large 
($\pm$0.5~dex) excursions in its assumed Sr content.
This occurs because the \ion{Sr}{2} line is as saturated in the secondary
as it is in the primary, but the central intensity of the unsmoothed 
spectrum is $\sim$0.2 of the continuum.  
Thus after the secondary's spectrum is diluted by the much larger flux of
the primary, a weak line that is relatively insensitive to abundance
changes ensues in the combined spectrum (and a naturally weaker line in the 
secondary simply becomes undetectable in the sum).

\begin{figure}
\epsscale{0.9}
\plotone{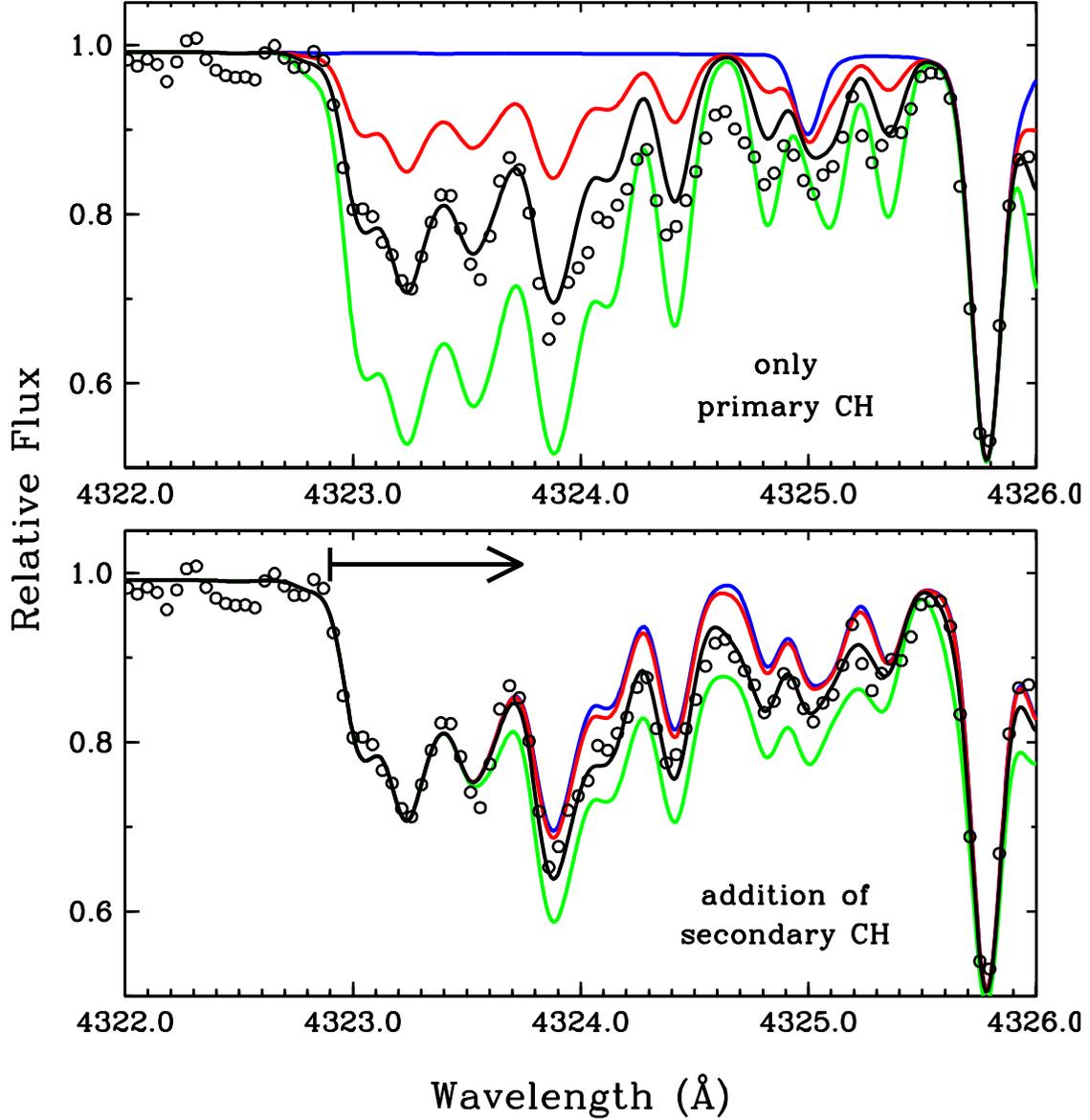}
\caption{
Observed and synthetic spectra of the CH G-bandhead in \cs22964.
The observation is 13817, for which $\Delta V_R$~= 58.09~\kmsec.
Open circles represent the observed spectrum.
In the top panel only the contribution of the primary star to this
blend is considered. 
The abundances of the synthetic spectra, in order of increasing CH
strength, are \eps{C}$_p$~= $-\infty$, 7.24, 7.64, and 8.04 (the blue,
red, black, and green lines, respectively).
In the bottom panel, the abundance of the primary is fixed at 
\eps{C}$_p$~= 7.64, and that of the secondary is, again in order of
increasing CH strength, \eps{C}$_s$~= $-\infty$, 6.44, 7.44, and 8.44.
An arrow indicates the wavelength offset between primary and secondary
stars of this observation.
\label{f7}}
\end{figure}

In Figure~\ref{f7} we illustrate the appearance of a small portion of
the CH G-band $A^{\rm 2}\Delta-X^{\rm 2}\Pi^+$ Q-branch bandhead in the 
observed combined-light spectra with a large velocity split.
This portion of the G-band has a sharp blue edge; the central wavelength 
of the first line is 4323.0~\AA. 
With $\Delta V_R$~= +58.1~\kmsec, the left edge of the secondary's 
bandhead begins at 4323.8~\AA, as indicated in the bottom panel of 
Figure~\ref{f7}.
In the top panel we show attempts to match the observed bandhead with
synthetic spectra that include only CH lines of the primary star.
When the spectral interval $\lambda$~$<$~4324.8~\AA\ is fit well, 
observed absorption is clearly missing at longer wavelengths in the 
synthetic spectrum.
This is completely solved by the addition of the secondary's CH bandhead,
at a comparable C abundance level to that of the primary, as shown in the 
bottom panel.

It is very difficult to derive reliable abundances in the \cs22964\
secondary even for the strong features illustrated here.
Nevertheless, it is clear that both components of this binary have
substantial overabundances of C and \ncap\ elements Sr and Ba.
Within the uncertainties of our analysis, the overabundance factors
for these elements appear to be the same.
Enhanced C accompanies \spro\ synthesis of \ncap\ elements during 
partial He-burning episodes of low/intermediate-mass stars, and the 
joint production of these elements is evident in the observed 
abundances of a number of BMP stars such as CS~29497-030 
(Sneden \etal\ 2003b\nocite{sne03b}; Ivans \etal\ 2005\nocite{iva05}).
However, a substantial C overabundance has also been seen in the
\rpro-rich star CS~22892-052 (Sneden \etal\ 2003a\nocite{sne03a}).
The Sr and Ba abundances determined to this point cannot distinguish between 
possible \ncap\ mechanisms that created the very heavy elements in \cs22964.

%%%%%%%%%%%%%%%%%%%%%%%%%%%%%%%%%%%
\section{ABUNDANCES FROM THE SYZYGY SPECTRUM\label{abtotal}}
%%%%%%%%%%%%%%%%%%%%%%%%%%%%%%%%%%%

\begin{figure}
\epsscale{0.9}
\plotone{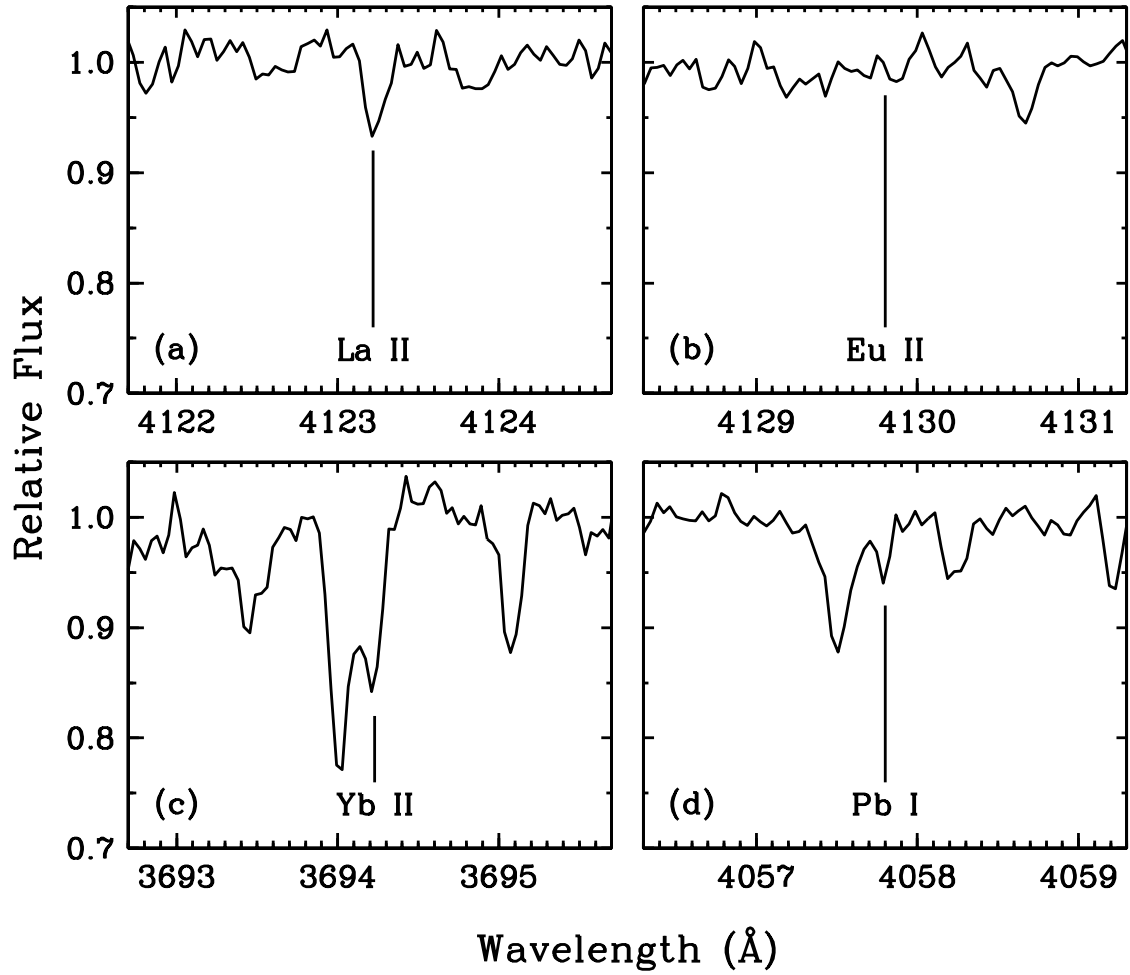}
\caption{
Selected spectral features of \ncap\ species in the \cs22964\ 
syzygy spectrum.
\label{f8}}
\end{figure}

To gain further insight into the \ncap\ element abundance distribution
we returned to the higher S/N mean syzygy spectrum.
Preston \etal\ (2006b)\nocite{pre06a} argued that the relative 
strengths of \ion{La}{2} and \ion{Eu}{2} lines can easily distinguish 
\spro\ dominance (stronger La features) from \rpro\ dominance (stronger 
Eu); see their Figure~1.
In panels (a) and (b) of Figure~\ref{f8} we show the same La and
Eu lines discussed by Preston \etal; the greater strength of the La feature
is apparent.

We computed a mean $EW$ from \ion{La}{2} lines at 3988.5, 3995.7,
4086.7, and 4123.2~\AA, and a mean $EW$ from \ion{Eu}{2} lines at 
3907.1, 4129.7, and 4205.1~\AA\ for a few warm metal-poor stars 
with \ncap\ overabundances.
The resulting ratio $<EW_{\rm La}>$/$<EW_{\rm Eu}>$~$\simeq$ 0.5 
for the \rpro-rich red horizontal-branch star CS~22886-043 
(Preston \etal\ 2006a)\nocite{pre06b}, $\simeq$~2.8 for the \spro-rich 
RR~Lyrae TY~Gruis (Preston \etal\ 2006a), and $\simeq$~2.7 for the 
$r$+$s$ BMP star CS~29497-030 (Ivans \etal\ 2005).\nocite{iva05}
For \cs22964\ we found $<EW_{\rm La}>$/$<EW_{\rm Eu}>$~$\simeq$~2.7, 
an unmistakable signature of an \spro\ abundance distribution.

For a more detailed \ncap\ element distribution for \cs22964\ we computed 
synthetic spectra of many transitions in the syzygy spectrum. 
We used the same binary synthesis version of the MOOG code that 
was employed for the velocity-split spectra shifted to 
$\Delta V_R$~=~0~\kmsec\ (as described in \S\ref{abboth}).
For these computations it was also necessary to assume that 
\eps{X}$_p$~=~\eps{X}$_s$ for all elements X.
We also derived abundances of a few lighter elements of interest
with this technique.
It should be noted that since the primary star is 4--6 times brighter than the 
secondary, abundances derived in this manner mostly apply to the primary.

In Table~\ref{tab5} we give the abundances for the \cs22964\ 
binary system.
For elements with abundances determined for the individual stars as 
discussed in \S\ref{abboth}, we also give estimates of their ``system''
abundances in this table.
These mean abundances were computed from the entries in 
Table~\ref{tab4}, giving both the abundances and their 
uncertainties ($\sigma$) of the primary star five times more 
weight than those of the secondary.
For elements with abundances derived from synthetic spectra of the syzygy
spectrum, the abundances and $\sigma$ values are means of the
results from individual lines, wherever possible.
For several of these elements only one transition was employed.
In these cases we adopted $\sigma$ = 0.20 or 0.25, depending on the
difficulties attendant in the synthetic/observed spectrum matches.
In the next few paragraphs we discuss the analyses of a few species 
that deserve special comment.

{\it Li I:} The resonance transition at 6707.8~\AA\ was easily 
detected in all \cs22964\ spectra, with $EW_{o,tot}$~$\simeq$~24.5~m\AA\ 
from the syzygy data. 
Our synthesis included only \iso{7}{Li}, but with its full 
hyperfine components.
The resolution and S/N combination of our spectra precluded any 
meaningful search for the presence of \iso{6}{Li}.
Reyniers \etal\ (2002)\nocite{rey02} have shown that the presence of a 
\ion{Ce}{2} transition at 6708.09~\AA\ can substantially contaminate 
the Li feature in \ncap-rich stars.
However, in our spectra the Ce line wavelength is too far from the 
observed feature, and our syntheses indicated that the Ce abundance 
would need to be about two orders of magnitude larger than our derived 
value (Table~\ref{tab5}) to produce measurable absorption in the 
\cs22964\ spectrum.

We attempted to detect the secondary \ion{Li}{1} line in two different ways.
First, we applied the subtraction technique (\S\ref{ewcalc}) to this
feature. 
However, the six velocity-split spectra yielded $EW_{o,p}$~=~23.5~m\AA\ 
with $\sigma$~=~3.5~m\AA\ (consistent with typical uncertainties for this
technique).
Thus the implied $EW_{o,s}$~=~1.0~m\AA\ is consistent with no detection
of the secondary's line.
Second, we co-added the velocity-split spectra after shifting them to the
rest system of the secondary.
A very weak ($EW_{o,s}$~$\simeq$~4~m\AA) line appears at 
$\lambda$~=~6707.9~\AA, about 0.1~\AA\ redward of the \ion{Li}{1} 
feature centroid.
We computed synthetic spectra assuming that the observed line is Li, and 
derived \eps{Li}~$\simeq$~+2.0 with an estimated uncertainty of $\pm$0.2
from the observed/synthetic fit.
This abundance is fortuitously close to the system value, but we do not
believe that the present data warrant a claim of Li detection in the
\cs22964\ secondary.

{\it Na I:} The 5682, 5688~\AA\ doublet, used in abundance analyses when the 
D-lines at 5890, 5896~\AA\ are too strong, is undetectably weak in \cs22964.
A synthetic spectrum match to the syzygy spectrum suggests that
\eps{Na}~$\lesssim$ 3.9.
The D~lines are strong, $\sim$40\% deep.  
Unfortunately, their profiles appear to be contaminated by telluric emission.
If we assume that the line centers and red profiles are unblended then
\eps{Na}~$\simeq$ 3.9.

Given the great strengths of the D-lines in the syzygy spectrum (dominated
by the primary star) we returned to the velocity-split spectra and 
searched for the secondary Na~D lines.
Absorptions a few percent deep at the expected wavelengths were seen
in all spectra.
Co-addition of these spectra after shifting to the rest velocity system of 
secondary yielded detection of both D-lines, and we estimated
\eps{Na}~$\simeq$ 3.8.
A problem with our analyses of the D-lines in all spectra was the lack of
telluric H$_{\rm 2}$O line cancellation; we did not observe suitable hot,
rapidly rotating stars for this purpose.
However, any unaccounted-for telluric features would drive the derived Na
abundances to larger values.  
Thus we feel confident that Na is not overabundant in \cs22964.

{\it CH:} Carbon was determined from CH G-band features in the 
wavelength range 4260--4330~\AA, with a large number of individual lines 
contributing to the average abundance.
Our CH line list was taken from the Kurucz (1998)\nocite{kur98} compendium.
The solar C abundance listed in Table~\ref{tab5} was determined with
the same line list (see Sneden \etal\ 2003b\nocite{sne03b}), rather than 
adopted from the recent revision of its abundance from other spectral
features by Allende Prieto, Lambert, \& Asplund (2002)\nocite{ale02}.

{\it Si I:} The only detectable line, at 3905.53~\AA, suffers potentially 
large  CH contamination, as pointed out by, \eg, Cohen \etal\ 
(2004)\nocite{coh04}.
These CH lines are very weak in ordinary warm metal-poor stars (Preston 
\etal\ 2006a)\nocite{pre06b}, but are strong in C-enhanced \cs22964.
We took full account of the CH in our synthesis of the \ion{Si}{1} line.
Note that the secondary's \ion{Si}{1} line can be detected in the 
velocity-split spectra, but it is always too blended with primary 
CH lines to permit a useful primary/secondary Si abundance analysis.

{\it Species with substructure:} Lines of \ion{Sc}{2}, \ion{Mn}{1}, 
\ion{Y}{2}, and \ion{La}{2} have significant hyperfine subcomponents,
which were explicitly accounted for in our syntheses (each of these 
elements have only one naturally-occurring isotope).
For \ion{Ba}{2} and \ion{Yb}{2} (whose single spectral feature at
3694.2~\AA\ is shown in panel (c) of Figure~\ref{f8}, 
both hyperfine and isotopic substructure were included.
We assumed an \spro\ distribution of their isotopic fractions: 
$f$(\iso{134}{Ba})~=~0.038, $f$(\iso{135}{Ba})~=~0.015, 
$f$(\iso{136}{Ba})~=~0.107, $f$(\iso{137}{Ba})~=~0.080, 
$f$(\iso{138}{Ba})~=~0.758; and
$f$(\iso{171}{Yb})~=~0.180, $f$(\iso{172}{Yb})~=~0.219, 
$f$(\iso{173}{Yb})~=~0.191, $f$(\iso{174}{Yb})~=~0.226, 
$f$(\iso{176}{Yb})~=~0.185.
We justify the \spro\ mixture choice below by showing that the total
abundance distribution, involving 17 elements, follows an \spro-dominant
pattern.

{\it Pb I:} The 4057.8~\AA\ line was easily detected in the syzygy spectrum 
(panel (d) of Figure~\ref{f8}), with $EW_{o,tot}$~$\sim$~5~m\AA.
This line is very weak, but we confirmed its existence and approximate
strength through co-addition of the six velocity-split spectra.
The Pb abundance was derived from synthetic spectra in which the
isotopic and hyperfine substructure were taken into account following
Aoki \etal\ (2002)\nocite{aok02}, and using the \ion{Pb}{1} line data
of their Table~4.
Variations in assumed Pb isotopic fractions produced no change in the
derived elemental abundance.
As with other features in the blue spectral region, CH contamination
exists, but our syntheses suggested that it is only a small fraction
of the \ion{Pb}{1} strength.

%%%%%%%%%%%%%%%%%%%%%%%%%%%%%%%%%%%%%%%%%%%%%%%%%%%%%%%%%%%%%%%%%%%%%
\section{IMPLICATIONS OF THE \cs22964\ ABUNDANCE PATTERN\label{history}}
%%%%%%%%%%%%%%%%%%%%%%%%%%%%%%%%%%%%%%%%%%%%%%%%%%%%%%%%%%%%%%%%%%%%%

\begin{figure}
\epsscale{0.9}
\plotone{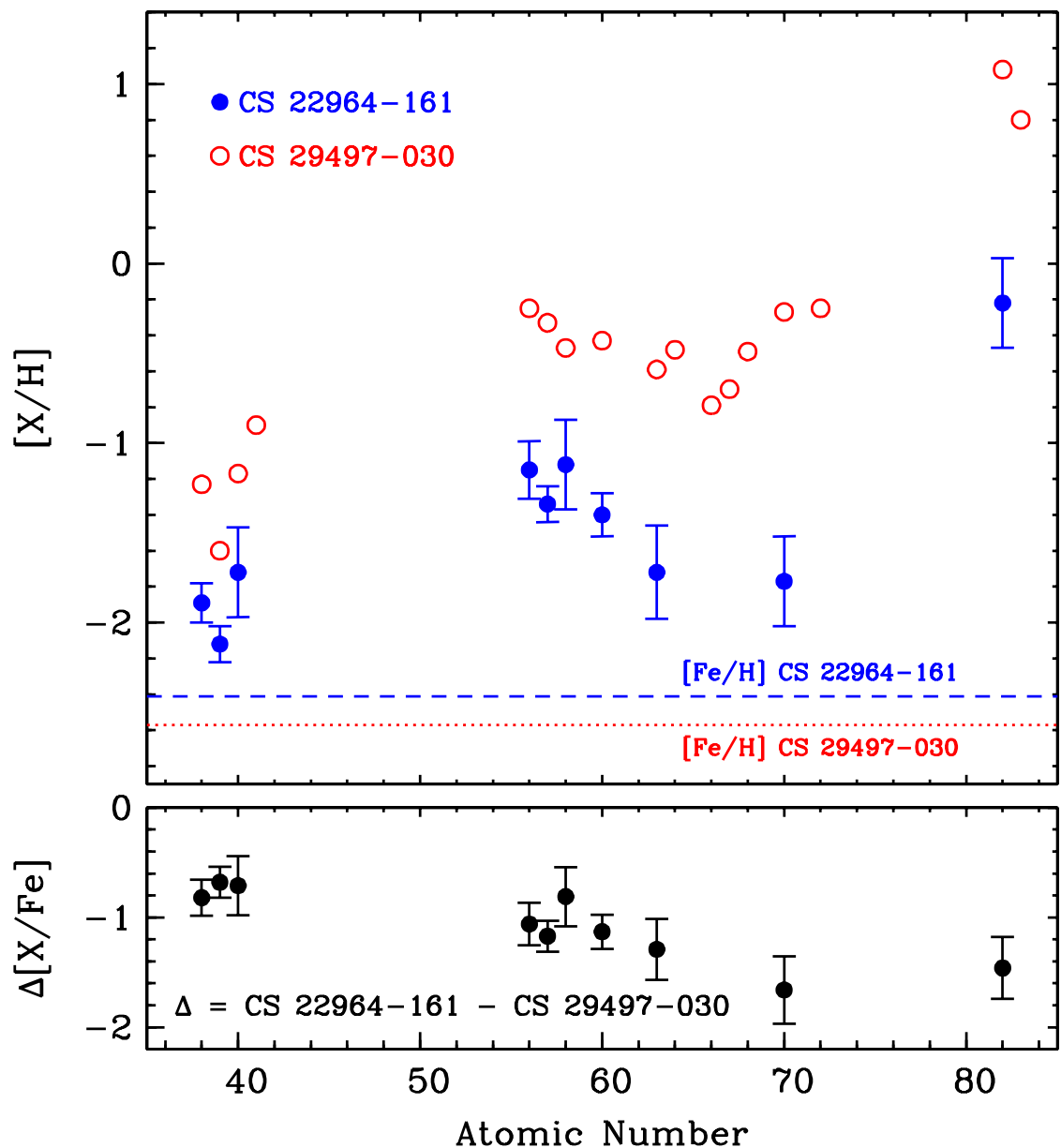}
\caption{
In the top panel, \ncap\ abundances relative to solar values [X/H] 
from the \cs22964\ combined spectrum are plotted along with those of 
the BMP very \spro-rich star CS~29497-030 (Ivans \etal\ 2005).
Horizontal lines are drawn to indicate the general Fe-metallicity levels 
of the two stars.
In the bottom panel the abundance differences between these two stars
are shown.
\label{f9}}
\end{figure}

In the top panel of Figure~\ref{f9} we plot the abundances relative
to solar values for the 10 \ncap\ elements detected in the \cs22964\ 
syzygy spectrum, together with the \ncap\ abundances in the very 
\spro-rich BMP star CS~29497-030.  
These two stars have nearly the same overall [Fe/H] metallicity, 
as indicated by the horizontal lines in the panel.
The \ncap\ overabundance pattern of \cs22964\ clearly identifies it
as another member of C- and \spro-rich ``lead star'' family.

However, the relative \ncap\ abundance enhancements [X/Fe] of CS~29497-030 
are about 1~dex higher than they are in \cs22964.
This is emphasized by taking the difference between the two abundance sets, 
as is displayed in the lower panel of Figure~\ref{f9}.
These differences illustrate the trend toward weaker \spro\ enhancements 
of the heaviest elements in \cs22964\ compared to CS~29497-030.

Many studies have argued that these overall abundance anomalies, 
now seen in many low metallicity stars, must have originated from mass 
transfer from a (former) AGB companion to the stars observed today.
Whole classes of stars with large C and \spro\ abundances are dominated by
single-line spectroscopic binaries, including the high-metallicity 
``Ba II stars'' (McClure, Fletcher, \& Nemec 1980\nocite{mcc80};
McWilliam 1988\nocite{mcw88}; McClure \& Woodsworth 1990\nocite{mcc90}), 
the low-metallicity red giant ``CH stars'' (McClure 1984\nocite{mcc84}),
the ``subgiant CH stars'' (McClure 1997)\nocite{mcc90}, and the
BMP stars (Preston \& Sneden 2000\nocite{pre00}; note that not all 
BMP stars share these abundance characteristics).
The case is especially strong for the subgiant CH and BMP stars, for these
are much too unevolved to have synthesized C and \spro\ elements in their
interiors and dredged these fusion products to the surfaces.
Our analysis has confirmed that primary and secondary \cs22964\ 
stars are on or near the main sequence. 
Therefore we suggest that a third, higher-mass star is now or 
once was a member of the \cs22964\ system.
During the AGB evolutionary phase of the third star, it transferred 
portions of its C- and \spro-rich envelope to the stars we now observe.

The very large Pb abundance of \cs22964\ strengthens the AGB 
nucleosynthesis argument.
Gallino \etal\ (1998\nocite{gal98}) and Travaglio \etal\ 
(2001\nocite{tra01}) predicted substantial Pb production
in \spro\ fusion zones of metal-poor AGB stars.
In metal-poor stars, the neutron-to-seed ratio is quite high, 
permitting the neutron-capture process to run through to Pb and Bi, 
the heaviest stable elements along the \spro\ path.  
Prior to this theoretical prediction, it was assumed that a $strong$ 
component of the \spro\ was required for the manufacture of half of the 
\iso{208}{Pb} in the solar system (Clayton \& Rassbach 1967\nocite{cr67}).
That the patterns of the abundances of the neutron-capture elements
in \cs22964\ and CS~29497-030 resemble each other so well is a
reflection of how easily low-metallicity AGB stars can synthesize
the heavy elements.

\subsection{Inferred Nucleosynthetic and Dilution 
                    History\label{nucleosynthesis}}

We explored the origin of the neutron-capture enhancements in \cs22964\ 
by comparing the derived photospheric abundances with predicted stellar 
yields from the \spro.
Employing FRANEC stellar evolutionary computations (see Straniero 
\etal\ 2003\nocite{straniero+03}; Straniero, Gallino, \& Cristallo 
2006\nocite{sgc06}), we performed nucleosynthetic calculations following 
those of Gallino \etal\  (1998\nocite{gal98}; 2006a\nocite{gal06a}) 
and Bisterzo \etal\ (2006\nocite{bisterzo+06}).  
We then sought out good matches between the observed and predicted abundance 
pattern distributions with low-mass AGB progenitors of comparable 
metallicities and a range of initial masses. 

We employed AGB models adopting different \iso{13}{C}-pocket efficiencies 
and  initial masses to explore the nucleosynthetic history of the 
observed chemical compositions (Busso, Gallino \& Wasserburg 
1999\nocite{busso+99}; Straniero \etal\ 2003\nocite{straniero+03}). 
Permitting some mixing of material between the \spro-rich 
contributions of the AGB donor, and the H-rich material of the 
atmosphere in which the contributions were deposited, we identified
those AGB model progenitors that would yield acceptable fits for the 
predicted yields of all \spro\ elements beyond Sr.

Ascertaining the best matches between the observed and predicted yields 
was performed in the following way.  
For a given initial AGB mass, we inspected the abundances of [Zr/Fe], 
[La/Fe], and [Pb/Fe] for various \iso{13}{C}-pocket efficiencies 
adopted in the calculations.  
The difference between our predicted [La/Fe] and that of the observed 
abundance ratio gave us a first guess to the dilution of material for a 
given \iso{13}{C}-pocket efficiency.
The dilution factor (dil) is defined as the logarithmic ratio
of the mass of the envelope of the observed star polluted with
AGB stellar winds and the AGB total mass transferred:
dil~$\equiv$~log$(M^{env}_{obs}/M^{AGB}_{transf})$.
The \iso{13}{C}-pocket efficiency is defined in terms of the 
\iso{13}{C} pocket mass involved in an AGB pulse that was adopted by 
Gallino \etal\ (1998)\nocite{gal98}.
Their \S2.2 states, ``The mass of the \iso{13}{C} pocket is 
5.0$\times$10$^{-4}$~\Msun\ [here called ST], about 1/20 of the typical 
mass involved in a thermal pulse. 
It contains 2.8$\times$10$^{-6}$~\Msun\ of \iso{13}{C}.''
Thus our shorthand notation for \iso{13}{C}-pocket efficiency will
be ST/N, where ''N'' is the reduction factor employed in generating
a particular set of AGB abundance predictions.
Subtracting this dilution amount from the predicted [Pb/Fe], we then 
compared the result to the observed [Pb/Fe].  
The large range of \iso{13}{C}-pocket efficiencies was then narrowed down, 
keeping only those results that fit the abundances of both [La/Fe] 
and [Pb/Fe] within 0.2 dex.  
Similar iterations were performed employing the abundances of [Zr/Fe].

\begin{figure}
\epsscale{0.75}
\plotone{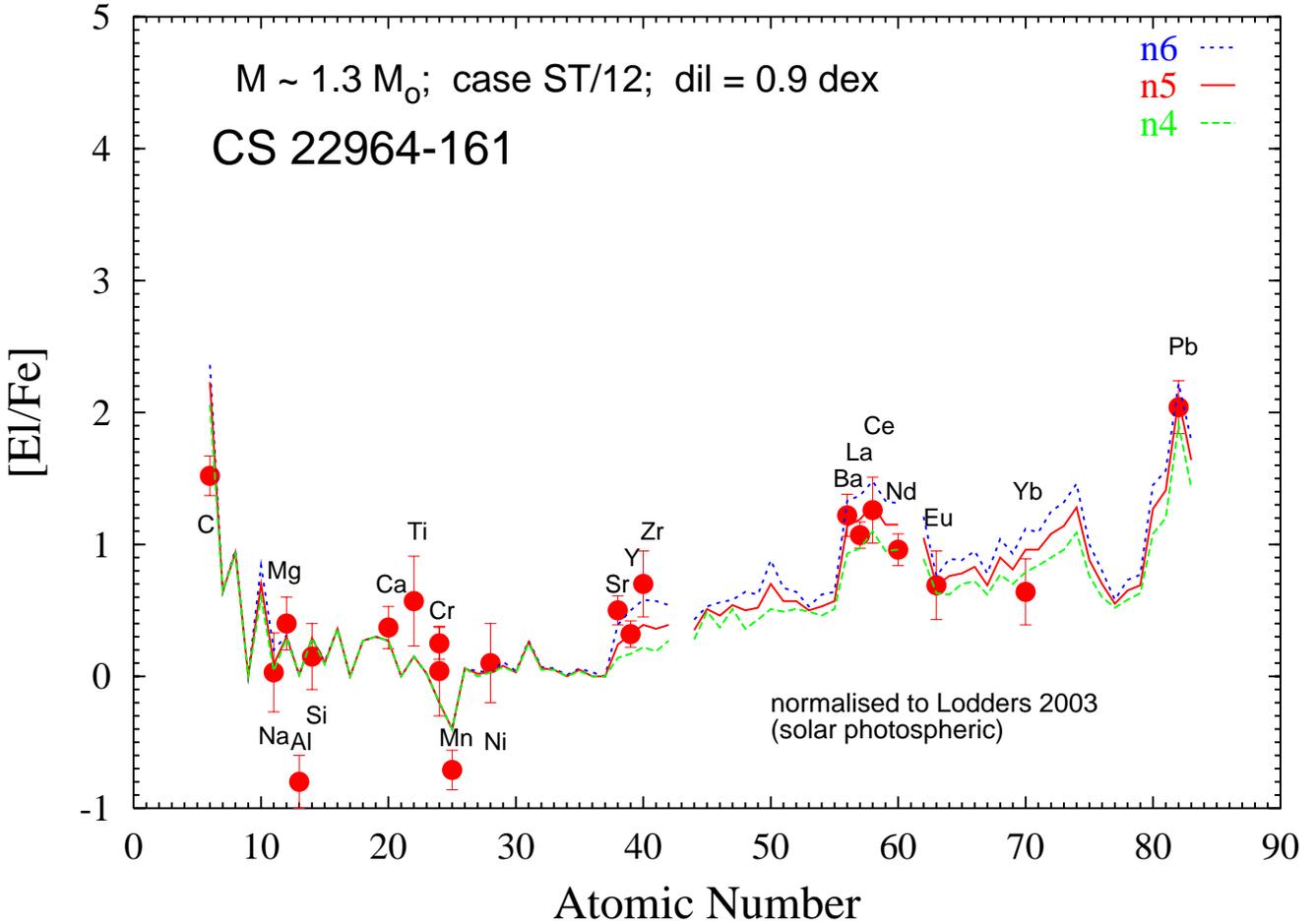}
\caption{Comparison of the [X/Fe] abundances in \cs22964\ with predictions 
from \spro\ calculations of a 1.3\Msun\ AGB star model.  
The solid red line corresponds to the best match between the observed 
and predicted abundance pattern.  
The dotted blue and dashed green lines indicate the difference that 
adopting models of $\pm$0.05\Msun\ would make (effectively, changing 
the number of thermal pulses from 4 to 6, identified in the figure
legend as n4, n5, n6).
For the specific choice of the \iso{13}{C}-pocket (ST/12), and for the
definition of the dilution factor ($dil$), see the text.
\label{f10}}
\end{figure}

\begin{figure}
\epsscale{0.75}
\plotone{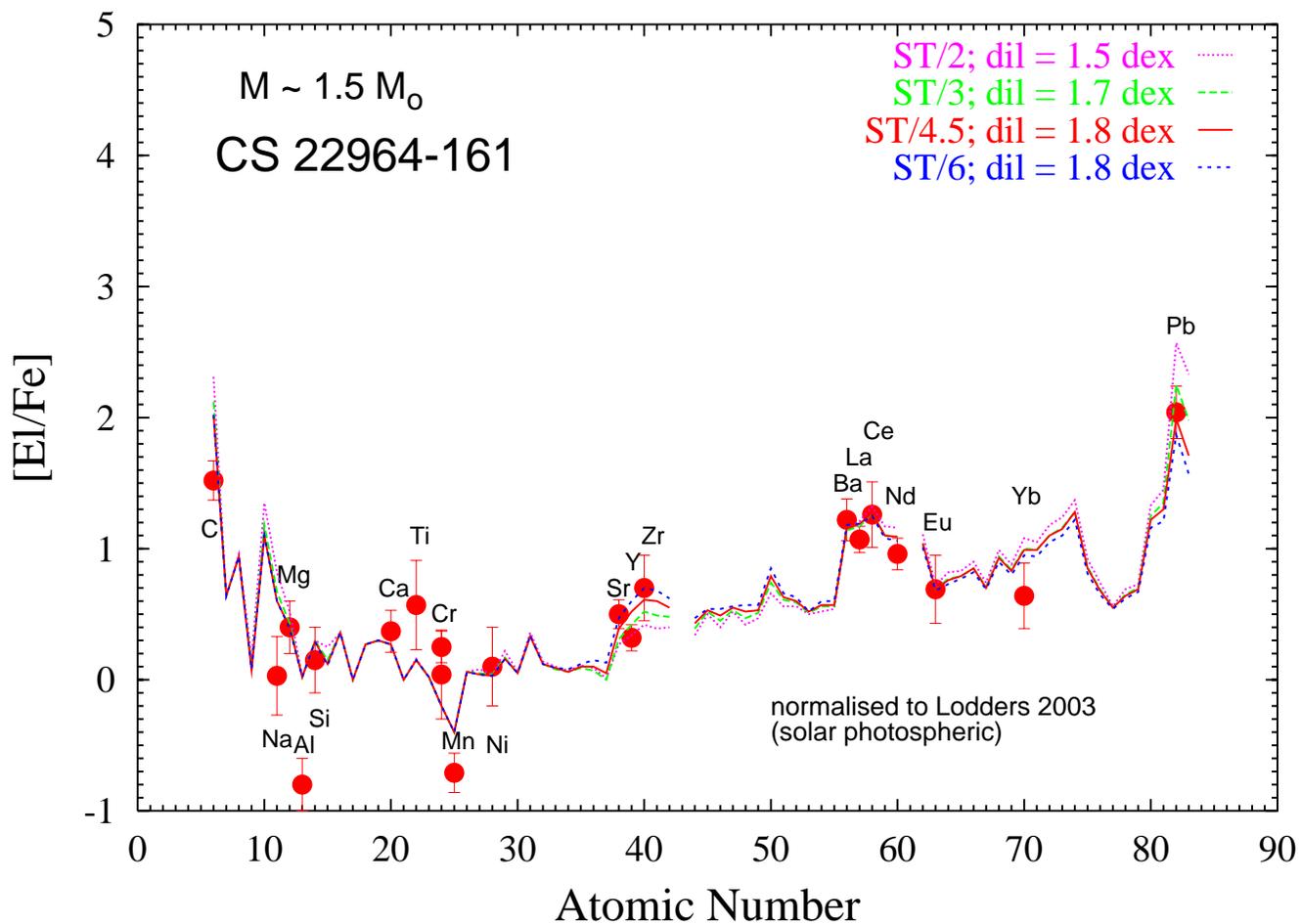}
\caption{Comparison of the [X/Fe] abundances in \cs22964\ with predictions 
from \spro\ calculations of a 1.5\Msun\ AGB star model.  
The solid red line corresponds to the best match between the observed 
and predicted abundance pattern.  
The dotted magenta, long-dashed green, and short-dashed blue lines show 
the difference made to the pattern by adopting different amounts 
of dilution.
For discussion of the the various choices of the \iso{13}{C}-pocket 
efficiency (ST/N), and the dilution factor (dil), see the text.
\label{f11}}
\end{figure}

We repeated this process for a range of initial AGB mass choices: 1.3, 
1.5, 2, 3, and 5\Msun.  
For AGB models of initial mass 3 and 5\Msun\ no good match was found 
because the light s-elements (ls~$\equiv$ Sr, Y, Zr) were predicted to
have too high abundances with respect to the heavy s-elements 
(hs~$\equiv$ Ba, La, Ce, Nd).
For AGB models of $M$~$\leq$~2\Msun\ a satisfactory solution was found for 
ls, hs, and Pb, provided a proper \iso{13}{C}-pocket efficiency and 
dilution factor were chosen.
In Figures~\ref{f10} and \ref{f11} we show 
the matches between predicted and observed abundances for the two
lowest-mass models, 1.3 and 1.5\Msun, respectively.

The abundances of the light elements Na and Mg further narrowed the
range of allowable AGB progenitor models.
Two independent channels are responsible for creating \iso{23}{Na}:
the \iso{22}{Ne}($p$,$\gamma$)\iso{23}{Na} reaction during H-shell burning,
and the neutron capture on \iso{22}{Ne} via the chain
\iso{22}{Ne}($n$,$\gamma$)\iso{23}{Ne}($\beta^-$$\nu$)\iso{23}{Na} in the
convective He flash (see Mowlavi 1999\nocite{mow99}; 
Gallino \etal\ 2006\nocite{gal06b}).
A large abundance of primary \iso{22}{Ne} derives from the primary 
\iso{12}{C} mixed with the envelope by previous third dredge up episodes, 
then converted to primary \iso{14}{N} by HCNO burning in the H-burning 
shell and followed by double $\alpha$ capture via the chain
\iso{14}{N}($\alpha$,$\gamma$)\iso{18}{F}($\beta^+$$\nu$)\iso{18}{O}($\alpha$,$\gamma$)\iso{22}{Ne}
during the early development of the convective thermal pulse. 
The next third dredge up episode mixes part of this primary 
\iso{22}{Ne} with the envelope. 
Finally, while the H-burning shell advances in mass, all the
\iso{22}{Ne} present in the H shell is converted to \iso{23}{Na} 
by proton capture, accumulating in the upper region of He intershell.
Note that in intermediate mass AGB stars suffering the so-called hot
bottom burning (HBB) in the deeper layers of their convective envelope,
efficient production of \iso{23}{Na} results from proton capture on 
\iso{22}{Ne} (Karakas \& Lattanzio 2003\nocite{kar03}). 
Furthermore, the marginal activation in the convective He flash of the 
reactions \iso{22}{Ne}($\alpha$,$n$)\iso{25}{Mg} and
\iso{22}{Ne}($\alpha$,$\gamma$)\iso{26}{Mg} would lead to enhanced Mg. 
At the same time, most neutrons released by  
\iso{22}{Ne}($\alpha$,$n$)\iso{25}{Mg} are captured by the very abundant 
primary \iso{22}{Ne}, thus producing \iso{23}{Na} through the second 
channel indicated above.

Comparison of the abundance data in Figures~\ref{f10} and 
\ref{f11} shows that the observed [Na/Fe] argues
for the exclusion of the 1.5\Msun\ AGB model, which over-predicts
this abundance ratio by $\sim$0.5~dex.
The same statement applies to the 2\Msun\ model, not illustrated here.
The larger Na abundances produced in these models, relative to the 
1.3\Msun\ model, is related to the larger number of thermal pulses 
(followed by third dredge up) that these stars experience.
These arguments leave the 1.3\Msun\ AGB model as the only one 
capable of providing a global good match to all observed elements.

In order to match these abundances, the predicted AGB yields 
require an inferred dilution by H-rich material.  
We do not know the distance between the AGB primary donor and the low 
mass companion (actually, the binary system) that is now observed.  
And, neither do we know precisely the mass loss rate of the AGB.  
Dilution by about 1~dex of \spro-rich AGB material with the original 
composition of the observed star, as we deduce on the basis of the above
nucleosynthesis analysis, only fixes the ratio of the accreted matter with 
the outer envelope of the observed star.  
A more detailed discussion is taken up in \S\ref{dilution}.

A final remark concerns C, for which Figures~\ref{f10} and
\ref{f11} indicate an $\sim$0.8~dex over-production
compared to the observed [C/Fe].
This mismatch can in principle be substantially reduced by the occurrence 
of the so-called cool bottom process (CBB: see Nollett, Busso, \& Wasserburg 
2003\nocite{nol03}; Dom\'inguez et al. 2004a,b\nocite{dom04a,dom04b}). 
This reduction would imply a substantial increase of predicted N abundance,
from [N/Fe]~$\sim$ +0.5 shown in Figures~\ref{f10} and
\ref{f11} to perhaps +1.5.
We attempted without success to detect the strongest CN absorption at the
3883~\AA\ bandhead.
Trial spectral syntheses of this wavelength region suggested that even
[N/Fe]~$\sim$ +1.5 would not produce detectable CN.

\subsection{Mixing and Dilution in the Stellar Outer Layers\label{dilution}}

Very recently Stancliffe \etal\ (2007)\nocite{sta07}, extending the 
earlier discussion of Theuns, Boffin, \& Jorissen (1996)\nocite{the96}, 
report that extensive mixing due to thermohaline instability in a 
metal-poor star on the main sequence may severely dilute material that 
has been accreted from a companion.
They computed the mixing time scale with the assumption that thermohaline 
convection behaved as a simple diffusion process, and they concluded 
that the new matter should rapidly be mixed down over about 90\% of the
stellar mass.
                                                                                
Treating thermohaline convection simply as diffusion does not take into 
account the special nature of this process, which has extensively been 
studied in oceanography (e.g. Veronis 1965\nocite{ver65}; 
Kato 1966\nocite{kat66}; Turner 1973\nocite{tur73}; Turner \& Veronis 
2000\nocite{tur00}; Gargett \& Ruddick 2003\nocite{gar03}), and has
also been considered in stars (Gough \& Toomre 1982\nocite{gou82}, 
Vauclair 2004\nocite{vau04}). 
Thermohaline convection occurs in the ocean when warm salty water comes 
on top of cool fresh water. 
In this case the stabilizing thermal gradient acts against the 
destabilizing salt gradient. 
If the stabilizing effect compensates the destabilizing one, the medium 
should be stable, but it remains unstable due to double-diffusion: 
when a blob begins to fall down, heat diffuses out of it more rapidly 
than salt; then the blob goes on falling as the two effects no longer 
compensate. 
This creates the well known ``salt fingers'', thus is a very different 
physical environment than ordinary convection.
                                                                                
A similar situation occurs in stars when high-$\mu$ matter comes upon a
lower-$\mu$ region, which is the case for hydrogen-poor accretion. 
This has been studied in detail for the case of planetary accretion on
solar-type stars (Vauclair 2004\nocite{vau04}). 
As shown in laboratory experiments (Gargett \& Ruddick 2003\nocite{gar03}) 
the fingers develop in a special layer with a depth related to the 
velocity of the blobs and to the dissipation induced by hydrodynamical 
instabilities at their edges. 
This is a complicated process which may also be perturbed by other 
competing hydrodynamical effects. 
Depending on the situation, it is possible that the ``finger'' regime stops 
before complete mixing of the high-$\mu$ material into the low-$\mu$ one. 
In this case, we expect that the final amount of matter that remains 
in the thin sub-photospheric convective zone depends only on the final 
$\mu$-gradient, whatever the original amount accreted: more accreted 
matter leads to more mixing so that the final result is the same. 
                                                                                
\begin{figure}
\epsscale{0.9}
\plotone{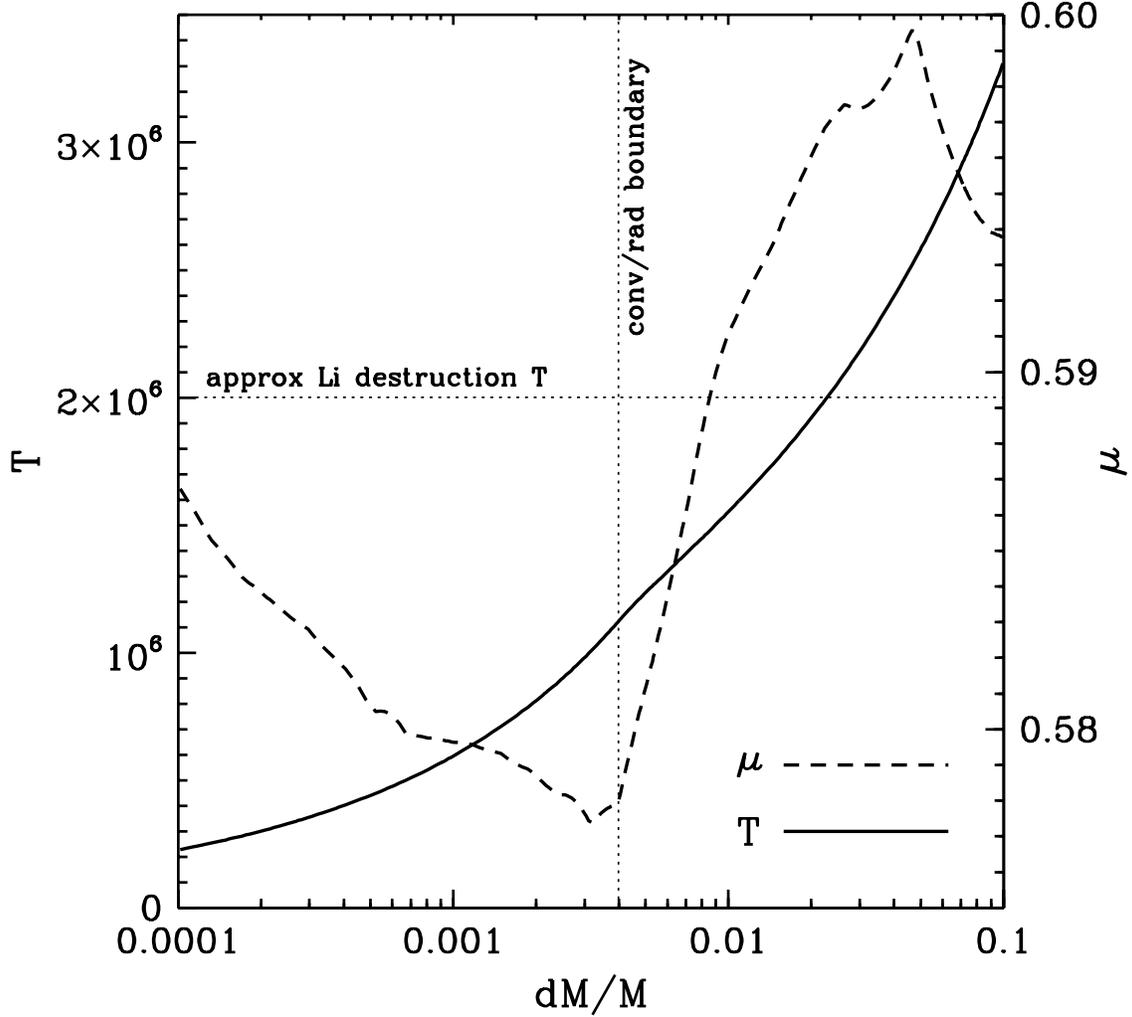}
\caption{Mean molecular weight ($\mu$) and temperature (T) profiles, as 
functions of the fractional mass (mass above the considered layer), in a 
0.78\Msun\ star with [Fe/H]~=~--2.3, at an age of 3.75~Gyr. 
Here $\mu$ is the real molecular weight, including partial ionization. 
The fractional mass of the convective zone is 0.004.  
The helium depletion due to gravitational settling, of about
20\% in this case, already leads to an important stabilizing
$\mu$-gradient below the convective zone. 
This model is one of those computed by Richard \etal\ (2002), including 
pure atomic diffusion. 
The full model was made available to us by Richard (private communication).
\label{f12}}
\end{figure}

Another very important point has to be mentioned. 
At the epoch when accretion occurs on the dwarfs, these stars are 
already about 3 to 4 Gyr old. 
Gravitational settling of helium and heavy elements has already
occurred and created a stabilizing $\mu$-gradient below the convective
zone (Vauclair 1999\nocite{vau99}; Richard, Michaud, \& Richer 
2002\nocite{ric02}). 
For example, as we illustrate in Figure~\ref{f12}, 
in a 0.78\Msun\ star with [Fe/H]~=~$-$2.3, after 3.75~Gyr 
the $\mu$-value is as low as 0.58 inside the convective zone while 
it goes up to nearly 0.60 in deeper layers. 
Thus the stabilizing $\Delta\mu / \mu$ is already of order 0.02.
If the star accretes hydrogen-poor matter, thermohaline convection may begin
but it remains confined by the $\mu$-barrier. The star can accrete as
much high-$\mu$ matter as possible until this stabilizing $\mu$-gradient
is flattened.
                                                                                
In \S\ref{nucleosynthesis} we saw, from the observed abundances and 
the AGB nucleosynthetic computations, that the dilution factor is of the 
order of 1~dex for the AGB model of initial mass $\sim$1.3\Msun.
Table~\ref{tab7} lists the predicted mass fraction of hydrogen ($X$),
helium ($Y$), and heavier elements ($Z$) in the envelope of AGB 
progenitor stars of 1.3 and 1.5\Msun\ for two cases.
The first case describes the final abundance mix produced by the AGB at 
the end of its nucleosynthetic lifetime (the last Third Dredge-Up).  
The second case describes the mass average of the winds from the AGB 
over its $s$-processing lifetime.  
Also noted is the mean molecular weight of the material ($\mu$), assuming 
fully ionized conditions.
From Table~\ref{tab7}, the $\mu$-value in the wind lies in the range 
0.61 to 0.78. 
From these computations, we find that, after dilution, the $\mu$-value 
inside the convective zone should increase by 
$\Delta\mu / \mu \simeq 0.1\%$ to $4\%$. 
Clearly most of the inferred values for the dilution factor are compatible 
with stellar physics. 
For the lowest values it is possible that the accreted matter simply 
compensates the effect of gravitational diffusion.
For larger values, thermohaline convection can have larger effects but, 
following Vauclair (2004\nocite{vau04}), if we accept an inverse 
$\mu$-gradient of order $\Delta\mu / \mu$ = 0.02 below the convective 
zone, such an accreted amount would still be possible.

\subsection{The Extraordinary Abundance of Lithium\label{lithium}}

Near main-sequence-turnoff CEMP stars display a variety of Li abundances.
The majority have undetectable 6708~\AA\ \ion{Li}{1} lines, implying
\eps{Li}~$<$~+1.
A few have abundances similar to the Spite \& Spite 
(1982)\nocite{spi82} ``plateau'': \eps{Li}~=~+2.10~$\pm$~0.09 
(Bonifacio \etal\ 2007)\nocite{bon07}.
These include our CEMP binary \cs22964, and at least CS~22898-027 
(Thorburn \& Beers 1992\nocite{tho92}) and LP~706-7 
(Norris, Ryan, \& Beers 1997\nocite{nor97}).
A few more stars have Li abundances somewhat less than the Spite plateau
value, \eg, \eps{Li}~$\simeq$~+1.7 in CS~31080-095 and CS~29528-041 
(Sivarani \etal\ 2006\nocite{siv06}).
This CEMP Li abundance variety stands in sharp contrast to the nearly 
constant value exhibited by almost all normal metal-poor stars near the 
main-sequence turnoff region.

Although \cs22964\ has a Spite plateau Li abundance, it is unlikely this
is the same Li with which the system was born.
Considering just the primary star, suppose that the C and \spro-rich 
material that it received was Li-free. 
If this material simply blanketed the surface of the primary,
mixing only with the original Li-rich atmosphere and outer 
envelope skin, then the resulting observed Li would be at least 
less than the plateau value.
Alternatively, if the transferred material induced more significant mixing 
of the stellar envelope, the surface Li would be severely diluted
below the limit of detectability in the primary's atmosphere.
In order to produce the large observed Li abundance in the primary,
freshly minted Li must have been transferred from the AGB donor star.

Li can be produced during the AGB phase both in massive AGB stars and
in low-mass AGB stars via
\iso{3}{He}($\alpha$,$\gamma$)\iso{7}{Be}($e^-,\nu_e$)\iso{7}{Li}, but hot 
protons rapidly destroy it via \iso{7}{Li}($p$,$\alpha$)\iso{4}{He}.
Cameron \& Fowler 1971\nocite{cam71}) proposed that in some circumstances
the freshly minted \iso{7}{Be} can however be transported to cooler interior
regions before decaying to \iso{7}{Li}.
In massive AGB stars ( 4\Msun~$\lesssim$~$M$~$\lesssim$~7\Msun) nucleosynthesis 
at the base of the convective envelope (``hot bottom burning''; Scalo, 
Despain, \& Ulrich 1975\nocite{sdu75}) can produce abundances via the
Cameron-Fowler mechanism as high as log~$\epsilon$(\iso{7}{Li})~$\sim$~4.5 
(Sackmann \& Boothroyd 1992\nocite{sb92}).  
However, the lifetimes of these massive AGB stars are far too short to be 
considered as a likely source of Li in \cs22964.

In models of low-mass AGB stars (1\Msun~$\leq$~$M$~$\le$~3\Msun) of 
[Fe/H]~=~$-$2.7, Iwamoto \etal\ (2004\nocite{iwamoto+04}) find that a 
H-flash episode can occur subsequent the first fully developed He shell flash.  
The convection produced by the He shell flash in this metal-poor star 
model can reach the bottom of the over-lying H-layer and bring protons down 
into the He intershell region, a region hot enough to induce a H flash.
The high-temperature conditions of the H-flash can then produce 
\iso{7}{Li} by the Cameron-Fowler mechanism.  
Although the Iwamoto \etal\ 2.5\Msun\ model produced a weak H
flash, the less massive models produced more energetic ones.
In the lowest mass models, surface abundances of
log~$\epsilon$(\iso{7}{Li})~$>$~3.2 were achieved.
Note that there is a maximum metallicity close to [Fe/H]~= $-$2.5 beyond 
which such an H-flash is not effective at producing \iso{7}{Li}.
Another viable mechanism for producing a very high abundance of \iso{7}{Li}
in low mass AGB stars of higher metallicities, as observed in the intrinsic 
C(N) star Draco~461 ([Fe/H]~= $-$2.0~$\pm$~0.2), was advanced by 
Dom\'inguez (2004a,b)\nocite{dom04a,dom04b} and is based on the operation
of the CBP introduced in \S\ref{nucleosynthesis}.
                                                                                
Although \spro\ calculations were not performed in the
Iwamoto \etal\ study, we note that it would seem that the
parameters of the best-fit model of \S\ref{nucleosynthesis} would
produce sufficient Li to fit the observations of \cs22964.

%%%%%%%%%%%%%%%%%%%%%%%%%%%%%%%%%%%
\section{IMPLICATIONS FOR THE PUTATIVE AGB DONOR STAR\label{agb}}
%%%%%%%%%%%%%%%%%%%%%%%%%%%%%%%%%%%

The binary nature of \cs22964\ presents a unique opportunity to explore
the nature of the AGB star presumably responsible for its abundance anomalies.
The ingredients of the exploration are:  {\it (a)} the analysis of our
radial velocities of \cs22964\ obtained over an interval of about 1100~days,
{\it (b)} estimation of the orbit changes caused by accretion from an
AGB companion, {\it (c)}  stability of the hierarchical triple system
in its initial and final states (Donnison \& Mikulskis 1992\nocite{don92},
hereafter DM92; Szebehely \& Zare 1977\nocite{sze77};
Tokovinin 1997\nocite{tok97}; Kiseleva, Eggleton, \& Anolova 
1994\nocite{kis94}), and {\it (d)} application of an appropriate 
initial-final mass relation for stars of intermediate mass (Weidemann 
2000\nocite{wei00}, and references therein).

\subsection{Evidence for an AGB relic?\label{relic}}
                                                                                
\begin{figure}
\epsscale{1.0}
\plotone{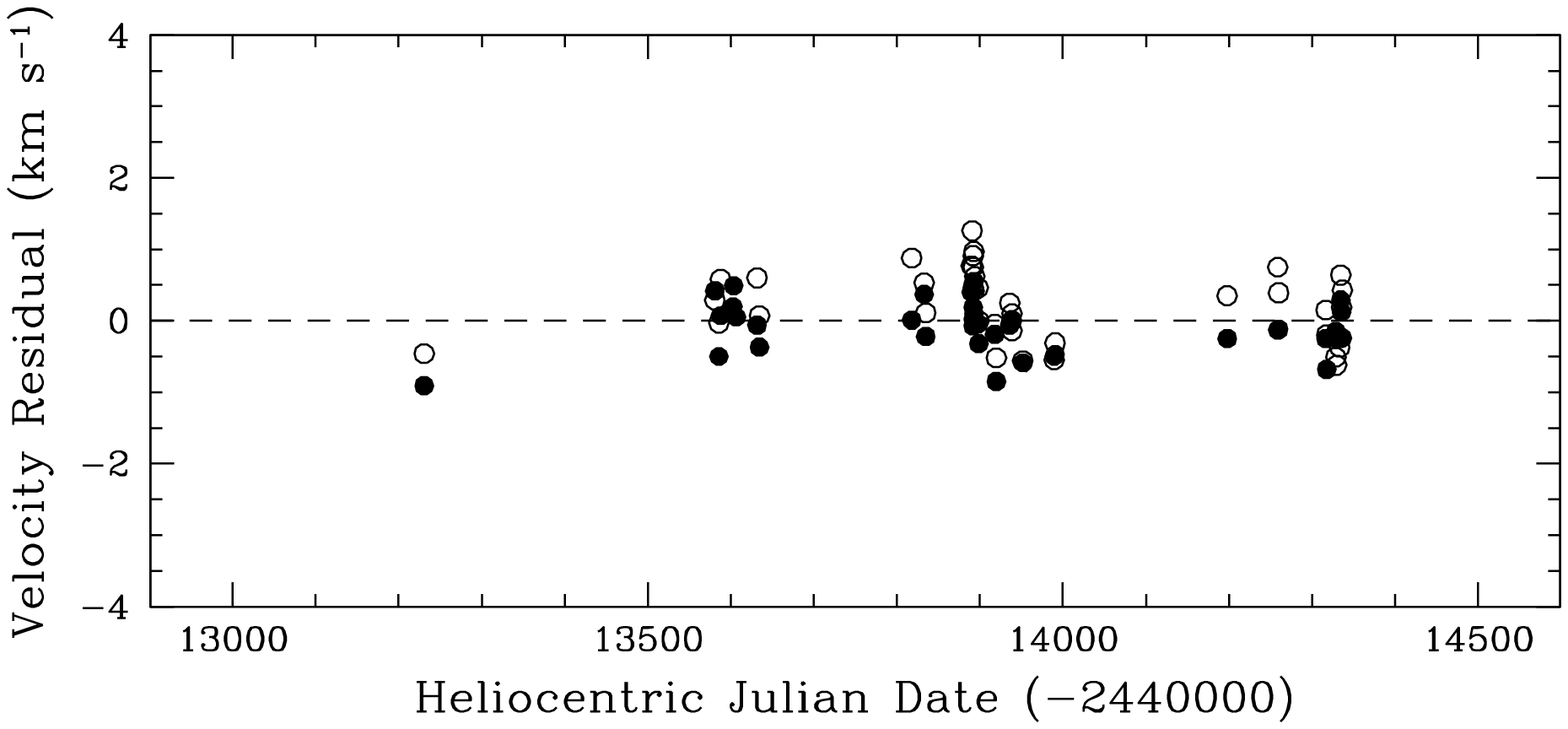}
\caption{
Velocity residuals to the fitted orbit are plotted against the date of 
observation for the Magellan data. 
Filled circles represent the primary and open circles represent the
secondary. 
These residuals show no trend with time.
\label{f13}}
\end{figure}

Velocity residuals calculated for our orbit solution show no trend
with time over the nearly 1100-day interval of observation.
This is illustrated by the plot of orbital velocity residuals versus
Julian Date in Figure~\ref{f13}.
Thus, we have yet to find evidence for a third component in the system.
The Julian Date interval of our \cs22964\ data is modest compared to
the longest orbital periods of the so-called Barium and CH giant stars
(McClure \& Woodsworth 1990)\nocite{mcc90} and main-sequence/subgiant
CH stars (McClure 1997)\nocite{mcc90} summarized in Table~\ref{tab6}.
On the other hand the errors of our observations ($\sigma$~=~0.5~\kmsec)
are small compared to typical velocity semi-amplitudes of these comparison
C-rich stellar samples, so drift of the center-of-mass velocity by as much
as 1 \kmsec\ due to the presence of a third star should be detectable.
Accordingly, Figure~\ref{f13} encourages us to contemplate the
possibility that \cs22964\ is not accompanied by a white dwarf relic of
AGB evolution.
Continuing observation will be required to test this notion.

\subsection{Change in Orbit Dimensions Induced by Accretion\label{orbchange}}
                                                                                
The consequences of mass accretion by a binary were first explored by
Huang (1956)\nocite{hua56}, who calculated the change in semi-major axis
of a binary embedded in a stationary interstellar cloud.
Using McCrea's (1953)\nocite{mcc53} ``retarding force'' that followed
from Bondi-Hoyle-Lyttleton accretion theory (Bondi \& Hoyle
1944)\nocite{bon44}, Huang found that mass accretion shrinks orbits, and
he proposed that this process could produce close binaries.
The problem of binary accretion has been revisited in recent years by
numerical simulations of star formation in molecular clouds.
Bate \& Bonnell (1997)\nocite{bat97} calculate the changes in binary
separation due to accretion by an infalling circumstellar cloud.
The change is very sensitive to the assumed specific angular momentum (SAM)
carried by the accreted material.
In the case of low SAM they recover Huang's result.
For large SAM orbits the binaries expand.
More recently Soker (2004)\nocite{sok04} has calculated the SAM of mass
accreted by a binary system embedded in the wind of an AGB star,
which is our case.
For triple systems containing a distant AGB companion with approximately
coplanar inner and outer orbital orbits (statistically most probable in the
absence of a special orientation mechanism) he also recovers Huang's result.
Finally, building on an earlier analysis by Davies \& Pringle
(1980)\nocite{dav80}, Livio \etal\ (1986)\nocite{liv86} found that the
rate of accretion of angular momentum from winds is significantly less
than that deposited in the simple Bondi-Hoyle accretion model.
                                                                                
Our gloss on all these analyses is that the most likely consequence of
mass transfer to \cs22964\ from its AGB companion a long time ago
was a decrease in separation, \ie, the initial separation was larger
than its present value of $\sim$0.9~AU.
Conservatively, we use 0.9 AU as a lower limit in the estimates that
follow below.
                                                                                
Empirical stability criteria for hierarchical triple systems are expressed
as the ratio of the semi-major axis of the outer orbit to
the semi-major axis of inner orbit (Heintz 1978\nocite{hei78};
Szebehely \& Zare 1977\nocite{sze77}), and alternatively as the ratio
of periods of the outer and inner orbits
(Kiseleva \etal\ 1994\nocite{kis94}; Tokovinin 1997\nocite{tok97}).
If these ratios exceed critical values, a triple system can be long-lived.
Note, however, the cautionary remarks of Szebehely \& Zare  about the
role of component masses in evaluation of stability for particular cases,
and the relevance of these masses to the conclusions of DM92 about 
the fate of our AGB relic.
                                                                                
Let $a_{12}$ be the semi-major axis of the relative orbit of \cs22964\
and $a_3$ be the relative semi-major axis that joins the center of mass
of \cs22964\ to the 3$^{rd}$ (AGB-to-be) star.
If we accept the Heintz (1978)\nocite{hei78} and Szebehely \& Zare
(1977)\nocite{sze77} stability criterion $a_3/a_{12}$~$>$~8, then
$a_3$~$>$~7.3~AU and the orbital period of the outer binary had to be
$P$~$>$~4000~d.
For Tokovinin's (1997)\nocite{tok97} more permissive criterion,
$P_3/P_{12}$~$>$~10 the minimum period is $P$~$>$~2500~d and
$a_3/a_{12}$~$>$~4.6.
Under these conditions none of our candidate metal-poor AGB stars (see the
discussion in \S\ref{nucleosynthesis}) with $M$~$<$~2\Msun\ fill the 
Roche lobes (Marigo \& Girardi 2007)\nocite{mar07} of these minimum orbits, 
so wind accretion must have been the mechanism of mass transfer.
According to Theuns \etal\ (1996)\nocite{the96} a 3\Msun\ AGB star in a
3-AU circular orbit will transfer no more than $\sim$1\% of its wind mass
to a 1.5 solar mass companion by wind accretion.
Applying their result to our case, $M$(AGB)~$\leq$~1.5\Msun\ and AGB relic
mass $\sim$0.5\Msun, the maximum mass transfer is about
(1.5$-$0.5)$\times$0.015~$\simeq$~0.015\Msun.
                                                                                
The remainder of the AGB wind mass is ejected from the system.
We calculated the change in period and semi-major axis of the AGB orbit
following Hilditch (2001)\nocite{hil01} for the case of a spherically
symmetric wind.
The orbit expands and the period lengthens according to
$${\rm ln}(a_f/a_i) = {\rm ln}(1+u_i) - {\rm ln}(1+u_f)$$  and
$${\rm ln}(P_f/P_i) = 2\times{\rm ln}(1+u_i) - 2\times{\rm ln}(1+u_f)$$
in which initial and final values of $u$ are
$u_i = M_{3i}/(M_1 + M_2)$ and $u_f = M_{3f}/(M_1 + M_2)$ .
We have ignored small changes in $M_1$ and $M_2$.
For $M_3$~=~1.3~\Msun\ and the Tokovinin (1997)\nocite{tok97} stability 
condition ($a_3/a_{12}$~$>$~4.6), the minimum orbit expands from 4.2~AU 
to 5.9~AU and the minimum period lengthens from 1900 days to 3700 days.
For the more restrictive Heintz ($a_3/a_{12}$~$>$~8) stability condition
the minimum orbit expands from 7.3~AU to 10.2~AU and the minimum period
lengthens from 4300 days to 8400 days.
The K-values for the center-of-mass of the \cs22964\ binary for the two
cases are 4.3~\kmsec\ and 3.3~\kmsec, respectively, both readily
detectable by conventional echelle spectroscopy.

\subsection{Dynamical Stability of \cs22964\ as a Hierarchical Triple System
             \label{dynamical}}
                                                                                
The DM92 study investigated the stability of triple systems that
consist of a close inner binary with masses ($M_1,M_2$) and semi-major
axis $a1$ attended by a remote companion of mass $M_3$ at
distance $q2$ from the center of mass of the inner binary.
They use the parameter notation of Harrington's (1975)\nocite{har75}
pioneering numerical simulations.
DM92 investigate stability in three mass regimes, by integration of the
equations of motion for various sets of masses, initial positions and
velocities: \\
\hspace*{1.5in} case {\it (i):}   $M_3$~$\leq$~minimum($M_1,M_2$) \\
\hspace*{1.5in} case {\it (ii):}  $M_1$~$\geq$~$M_3$~$\geq$~$M_2$ \\
\hspace*{1.5in} case {\it (iii):} $M_3$~$\geq$~maximum($M_1,M_2$) \\
They pursued the integrations for at least 1000 orbits of the inner
binary or until disruption, which ever happened first.
Among the several results in \S5 of their paper one is of particular
interest for us: in their words ``When $M_3$ was the least massive body in
the system, the system invariably became unstable through the tendency
of $M_3$ to escape from the system altogether.''
Accepting the McClure (1984)\nocite{mcc84} paradigm, we suppose that
\cs22964\ began its existence as a case {\it (iii)} hierarchical triple
system in which a relatively low-mass AGB progenitor 
(\S\ref{nucleosynthesis}) was orbited by a close binary with component 
masses only slightly smaller than their present values, 
$M_1$sin$^3i$~=~0.77\Msun\ and $M_2$sin$^3i$~=~0.68~\Msun.
Evidently, the distance between $M_3$ and the close binary was
large enough to insure stability during the lifetime of $M_3$.
Curiously, although Kiseleva \etal\ (1994)\nocite{kis94} made numerical
simulations apparently similar to those of DM92, they make no mention of the
ejection of low-mass outer components that attracted our attention to DM92.
                                                                                
We confine our attention to initial values of AGB mass $<$2.0\Msun,
adopted in accordance with the nucleosynthetic calculations in 
\S\ref{nucleosynthesis}.
In the late stages of AGB evolution the system evolved from case
{\it (iii)} to case {\it (i)} via mass loss due to the AGB superwind.
For all acceptable models in \S\ref{nucleosynthesis} the mass of the WD 
relic for an AGB initial mass model of 1.3\Msun\ is less than 
0.60\Msun\ (Weidemann 2000\nocite{wei00}).
Therefore, according to DM92 $M_3$ will be ejected following this 
ancient AGB evolution.
Our extant radial velocity data suggest that the ejection already occurred.
From the effect of the AGB relic on the $\gamma$ velocity of CS 22964-161 
calculated above we believe that this conclusion can be tested by future 
observations.

%%%%%%%%%%%%%%%%%%%%%%%%%%%%%%%%%%%
\section{CONCLUSIONS}
%%%%%%%%%%%%%%%%%%%%%%%%%%%%%%%%%%%

We have obtained high resolution, high S/N spectra over a three-year
period for \cs22964, which was known to be metal-poor and C-rich from 
previous low resolution spectroscopic studies.
We discovered \cs22964\ to be a double-line spectroscopic binary, and
have derived the binary orbital parameters ($P$~=~252~d, $e$~=0.66, 
$M_{p}~sin^{3}~i$~= 0.77~\Msun, $M_{s}~sin^{3}~i$~= 0.68~\Msun).
Both binary members lie near the metal-poor main-sequence turnoff
(\teff$_{,p}$~= 6050~K, \logg$_p$~= 3.7, and \teff$_{,s}$~=5850~K,
\logg$_s$~= 4.1).
We derived similar overall metallicities for primary and secondary,
[Fe/H]~= $-$2.4 and similar abundance ratios [X/Fe].
In particular, both stars are similarly enriched in C and the \ncap\
elements, with a clear \spro\ nucleosynthesis signature.
The primary has a large Li content; the secondary's \ion{Li}{1} feature
is undetectably weak, and does not usefully constrain its Li abundance.

The observed Li, C and \spro\ abundances of the \cs22964\ system must have 
been produced by an AGB star whose relic, depending on its mass, may have 
been ejected from the post-AGB hierarchical triple.  
It seems AGB enrichment was produced by minor mass accretion in the AGB
super-wind rather than Roche-lobe overflow. 
As discussed in \S\ref{history}, and contrary to the thermohaline diffusion
suggestion of Stancliffe \etal\ (2007\nocite{sta07}), such hydrogen-poor 
accreted matter probably remained in the outer stellar layers, owing to the
stabilizing $\mu$-gradient induced by helium gravitational settling.
Strong thermohaline diffusion is difficult to reconcile with the observed
Li:  about 9/10 of the stellar mass would be efficiently mixed on a time 
scale of 1~Gyr, which would imply the complete destruction of \iso{7}{Li} 
because of the very high temperature reached in the inner zones.
It seems more likely that the moderate thermohaline mixing discussed 
\S\ref{history}, related to the effect of gravitational settling in the 
first 3 to 4~Gyr before mass accretion by the AGB donor, would save the 
Li from destruction. 
In the model of a low mass AGB of low metallicity calculated by 
Dom\'inguez \etal\ (2004a,b\nocite{dom04a,dom04b}) with the inclusion of 
``cool bottom processing'' during the AGB phase, \eps{Li}~$>$ 3 is achieved
in the envelope.
A dilution factor of about 1~dex would bring the Li abundance into accord
with the observed value.

%%%%%%%%%%%%%%%%%%%%%%%%%%%%%%%%%%%
\acknowledgments
%%%%%%%%%%%%%%%%%%%%%%%%%%%%%%%%%%%
It is a pleasure to thank Guillermo Torres for sharing his TODCOR 
and orbit solution source code with us, Oscar Straniero for helpful 
discussions and continuous advice on AGB models of low-mass and very low
metallicity,  and Noam Soker 
for helpful discussions on wind accretion in binary stars.
We are grateful to Olivier Richard for kindly providing one of his 
population~II stars models, including consistently computed atomic diffusion.
We thank Alan Boss, Janusz Kaluzny, Wojtek Krzeminski, and
Kamil Zloczewski for obtaining photometric observations.
Some of this work was accomplished while CS was a Visiting Scientist 
at the Carnegie Observatories. The hospitality and support of the 
observatory director is greatly appreciated.  
This work has been supported by the U.S. National Science Foundation 
through grants AST-0307495 and AST-0607708 to CS and AST-0507325 to IBT,
and by by the Italian MIUR-PRIN06 Project ``Late phases of Stellar 
Evolution: Nucleosynthesis in Supernovae, AGB stars, Planetary Nebulae''
to RG.

%%%%%%%%%%%%%%%%%%%%%%%%%%%%%%%%%%%
%      HERE ARE REFERENCES
%%%%%%%%%%%%%%%%%%%%%%%%%%%%%%%%%%%

%%%%%%%%%%%%%%%%%%%%%%%%%%%%%%%%%%%
%      HERE ARE FIGURES
%%%%%%%%%%%%%%%%%%%%%%%%%%%%%%%%%%%
\clearpage

\clearpage
                                                                                
\clearpage
                                                                                
\clearpage

\clearpage

\clearpage
                                                                                
\clearpage
                                                                                
\clearpage

\clearpage

\clearpage

\clearpage

\clearpage
                                                                                
\clearpage

%%%%%%%%%%%%%%%%%%%%%%%%%%%%%%%%%%%
%      HERE ARE TABLES
%%%%%%%%%%%%%%%%%%%%%%%%%%%%%%%%%%%

\clearpage

\label{data}

\begin{deluxetable}{lrrr}
\tablecolumns{4}
\tablewidth{0pt}
\tabletypesize{\normalsize}
\tablecaption{Velocity Measurements for CS22964-161\label{tab1}}
\tablehead{
\colhead{HJD}         &
\colhead{V$_{p}$}     &
\colhead{V$_{s}$}     &
\colhead{Phase}       \\
\colhead{(- 2440000)} &
\colhead{km s$^{-1}$} &
\colhead{km s$^{-1}$} &
\colhead{}       
}

\startdata  
\multicolumn{4}{c}{du Pont Observations} \\ \hline 
  13592.52532 &  24.79 &  41.04 &  0.0716 \\
  13592.56468 &  24.63 &  42.35 &  0.0718 \\
  13593.52234  & 25.34 &  38.32 &  0.0756 \\
  13593.56091 &  25.21 &  40.26 &  0.0757 \\
  13867.92873 &  36.59 &  27.00 &  0.1623 \\
  13870.91114 &  36.44 &  25.90 &  0.1741 \\
  13871.85817 &  37.22 &  26.27 &  0.1778 \\ \hline
\multicolumn{4}{c}{ Magellan Observations} \\ \hline
  13230.66465 &  38.43 &  25.01 &  0.6386 \\
  13580.79663  &  5.09 &  65.21 &  0.0252 \\
  13585.74591 &  14.80 &  52.79 &  0.0448 \\
  13587.62635 &  18.56 &  49.77 &  0.0522 \\
  13602.62780 &  32.29 &  \nodata &  0.1116 \\
  13603.50474 &  33.00 &  \nodata &  0.1151 \\
  13606.50536 &  33.83 &  \nodata &  0.1270 \\
  13631.67197 &  39.08 &  26.29 &  0.2266 \\
  13634.54878 &  39.06 &  25.43 &  0.2380 \\
  13817.90625 &   6.10 &  64.19 &  0.9642 \\
  13832.88449 &   4.11 &  66.51 &  0.0235 \\
  13834.88836 &   8.21 &  60.76 &  0.0314 \\
  13889.88563 &  40.09 &  25.83 &  0.2492 \\
  13890.73000 &  40.19 &  26.24 &  0.2526 \\
  13891.72338 &  40.03 &  25.80 &  0.2565 \\
  13891.79617 &  39.87 &  25.63 &  0.2568 \\
  13891.91199 &  39.79 &  25.33 &  0.2572 \\
  13892.72313 &  40.47 &  25.77 &  0.2604 \\
  13892.91348 &  40.03 &  25.21 &  0.2612 \\
  13893.76837 &  40.02 &  25.34 &  0.2646 \\
  13897.80466 &  40.22 &  24.87 &  0.2806 \\
  13898.72511 &  40.00 &  24.35 &  0.2842 \\
  13917.74748 &  40.82 &  23.51 &  0.3595 \\
  13919.72723 &  40.20 &  23.00 &  0.3674 \\
  13935.76042 &  41.08 &  23.66 &  0.4309 \\
  13938.71168 &  41.14 &  23.29 &  0.4426 \\
  13938.73591 &  41.10 &  23.53 &  0.4427 \\
  13951.57253 &  40.34 &  23.09 &  0.4935 \\
  13989.60930 &  38.73 &  25.04 &  0.6441 \\
  13990.59320 &  38.69 &  25.36 &  0.6480 \\
  14197.85590 &  40.80 &  23.87 &  0.4689 \\
  14258.82674 &  37.60 &  28.04 &  0.7088 \\
  14259.78553 &  37.50 &  27.81 &  0.7126 \\
  14316.78097 &  15.19 &  52.81 &  0.9383 \\
  14317.65762 &  13.63 &  53.75 &  0.9418 \\
  14328.67079 &  -4.33 &  74.49 &  0.9854 \\
  14329.70221 &  -5.60 &  75.71 &  0.9895 \\
  14333.56647 &  -5.03 &  75.31 &  0.0048 \\
  14334.49082 &  -3.13 &  74.78 &  0.0085 \\
  14335.48607 &  -1.45 &  72.22 &  0.0124 \\
  14336.53932 &   0.46 &  69.88 &  0.0166 \\
\enddata

\end{deluxetable}

\clearpage

\begin{deluxetable}{lr}
\tablecolumns{2}
\tablewidth{0pt}
\tabletypesize{\normalsize}
\tablecaption{Orbital Parameters for CS22964-161\label{tab2}}
\tablehead{
\colhead{Parameter}       &
\colhead{Value}             
}

\startdata
$P$ (days)                       & 252.481$\pm$0.043 \\ 
$T_{0}$~(HJD-244 0000)           & 13827.383$\pm$0.103 \\
$\gamma~(km~s^{-1})$             & 32.85$\pm$ 0.05 \\
$K_{p}~(km~s^{-1})$              & 23.65$\pm$0.10 \\
$K_{s}~(km~s^{-1})$              & 26.92$\pm$0.13 \\
$e$                              & 0.6564$\pm$0.0024 \\
$\omega$ (deg)                   & 188.54$\pm$0.39 \\
$\sigma_{p}~(km~s^{-1})$         & 0.37 \\
$\sigma_{s}~(km~s^{-1})$         & 0.56 \\
Derived quantities: & \\
$A~sin~i$~($R_{\odot}$)          & 190.32$\pm$0.65 \\
$M_{p}~sin^{3}~i$~($M_{\odot}$)  & 0.773$\pm$0.009 \\
$M_{s}~sin^{3}~i~$($M_{\odot}$)  & 0.680$\pm$0.007 \\
\enddata

\end{deluxetable}

\clearpage

\begin{deluxetable}{lrrrrrrr}
\tablecolumns{8}
\tablewidth{0pt}
\tabletypesize{\tiny}
\tablecaption{Line List for CS~22964$-$161\label{tab3}}
\tablehead{
\colhead{Species}                       &
\colhead{$\lambda$}                     &
\colhead{$\chi$}                        &
\colhead{log $gf$}                      &
\colhead{$EW_{p,o}$}                    &
\colhead{$EW_{p,t}$}                    &
\colhead{$EW_{s,o}$}                    &
\colhead{$EW_{s,t}$}                    \\
\colhead{}                              &
\colhead{\AA}                           &
\colhead{eV}                            &
\colhead{}                              &
\colhead{m\AA}                          &
\colhead{m\AA}                          &
\colhead{m\AA}                          &
\colhead{m\AA}                          
}

\startdata
Li I  & 6707.80 & 0.00 & $+$0.18 &      24 &      29 & \nodata & \nodata \\
Mg I  & 3829.36 & 2.71 & $-$0.21 &     104 &     122 &      31 &     222 \\
Mg I  & 3832.31 & 2.71 & $+$0.14 &     128 &     150 &      30 &     213 \\
Mg I  & 4703.00 & 4.35 & $-$0.38 &      40 &      47 &       8 &      55 \\
Mg I  & 5172.70 & 2.71 & $-$0.38 &     127 &     152 &      25 &     156 \\
Mg I  & 5183.62 & 2.72 & $-$0.16 &     144 &     172 &      28 &     174 \\
Mg I  & 5528.42 & 4.35 & $-$0.34 &      40 &      49 &      10 &      63 \\
Al I  & 3961.53 & 0.01 & $-$0.34 &      62 &      73 &      11 &      79 \\
Si I  & 3905.53 & 1.91 & $+$0.09 & \nodata & \nodata & \nodata & \nodata \\
Ca I  & 4226.74 & 0.00 & $+$0.24 &     134 &     158 &      27 &     180 \\
Ca I  & 5349.47 & 2.71 & $-$0.31 &      11 &      13 & \nodata & \nodata \\
Ca I  & 5588.76 & 2.53 & $+$0.36 &      21 &      28 &       3 &      27 \\
Ca I  & 5857.46 & 2.93 & $+$0.24 &      17 &      21 & \nodata & \nodata \\
Ca I  & 6102.73 & 1.88 & $-$0.79 &      12 &      14 &       4 &      23 \\
Ca I  & 6122.23 & 1.89 & $-$0.32 &      26 &      33 &       2 &      12 \\
Ca I  & 6162.18 & 1.90 & $-$0.09 &      33 &      41 &       4 &      27 \\
Ca I  & 6439.08 & 2.52 & $+$0.39 &      28 &      36 &       3 &      22 \\
Sc II & 3645.31 & 0.02 & $-$0.42 &      43 &      50 &       8 &      61 \\
Sc II & 4246.84 & 0.31 & $+$0.24 &      59 &      69 &       8 &      55 \\
Ti II & 3477.19 & 0.12 & $-$1.06 &      85 &      99 & \nodata & \nodata \\
Ti II & 3510.85 & 1.89 & $+$0.14 &      55 &      64 & \nodata & \nodata \\
Ti II & 3596.05 & 0.61 & $-$1.22 &      43 &      50 &      12 &      90 \\
Ti II & 3641.34 & 1.24 & $-$0.71 &      40 &      46 &      12 &      86 \\
Ti II & 3759.30 & 0.61 & $+$0.20 &      97 &     113 &      19 &     137 \\
Ti II & 4394.06 & 1.22 & $-$1.77 &      16 &      18 & \nodata & \nodata \\
Ti II & 4443.80 & 1.08 & $-$0.70 &      55 &      66 &       6 &      45 \\
Ti II & 4444.56 & 1.12 & $-$2.21 &      14 &      16 & \nodata & \nodata \\
Ti II & 4468.49 & 1.13 & $-$0.60 &      59 &      71 &       7 &      52 \\
Ti II & 4501.27 & 1.12 & $-$0.76 &      54 &      65 &       5 &      40 \\
Ti II & 5336.79 & 1.58 & $-$1.63 &      13 &      16 & \nodata & \nodata \\
Cr I  & 3578.69 & 0.00 & $+$0.41 &      59 &      69 &      13 &      97 \\
Cr I  & 3593.50 & 0.00 & $+$0.31 &      65 &      76 &       6 &      46 \\
Cr I  & 4254.33 & 0.00 & $-$0.11 &      57 &      67 &       4 &      29 \\
Cr I  & 4274.80 & 0.00 & $-$0.23 &      72 &      84 & \nodata & \nodata \\
Cr I  & 4289.72 & 0.00 & $-$0.36 &      63 &      74 &       7 &      44 \\
Cr I  & 4646.17 & 1.03 & $-$0.73 &       5 &       6 &       2 &      14 \\
Cr I  & 5206.04 & 0.94 & $+$0.03 &      29 &      35 &       6 &      36 \\
Cr I  & 5409.79 & 1.03 & $-$0.71 &      15 &      18 & \nodata & \nodata \\
Cr II & 4558.65 & 4.07 & $-$0.66 &       8 &       9 &       2 &      10 \\
Cr II & 4588.20 & 4.07 & $-$0.64 &       6 &       7 & \nodata & \nodata \\
Mn I  & 4030.76 & 0.00 & $-$0.48 & \nodata & \nodata & \nodata & \nodata \\
Mn I  & 4033.06 & 0.00 & $-$0.62 & \nodata & \nodata & \nodata & \nodata \\
Mn I  & 4034.49 & 0.00 & $-$0.81 & \nodata & \nodata & \nodata & \nodata \\
Fe I  & 3475.46 & 0.09 & $-$1.05 &      82 &      94 &      34 &     249 \\
Fe I  & 3476.71 & 0.12 & $-$1.51 &      73 &      84 &       8 &      61 \\
Fe I  & 3490.59 & 0.05 & $-$1.11 &      76 &      88 &      16 &     117 \\
Fe I  & 3497.84 & 0.11 & $-$1.55 &      71 &      82 &       8 &      58 \\
Fe I  & 3521.27 & 0.91 & $-$0.99 &      56 &      65 &      15 &     111 \\
Fe I  & 3554.94 & 2.83 & $+$0.54 &      47 &      55 &       6 &      45 \\
Fe I  & 3558.53 & 0.99 & $-$0.63 &      68 &      79 &      16 &     113 \\
Fe I  & 3565.40 & 0.96 & $-$0.13 &      89 &     103 &      21 &     151 \\
Fe I  & 3570.13 & 0.91 & $+$0.18 &     151 &     175 &      26 &     190 \\
Fe I  & 3581.21 & 0.86 & $+$0.42 &     114 &     132 &      36 &     262 \\
Fe I  & 3606.69 & 2.69 & $+$0.32 &      41 &      47 &       8 &      59 \\
Fe I  & 3608.87 & 1.01 & $-$0.09 &      80 &      92 &      20 &     142 \\
Fe I  & 3618.78 & 0.99 & $+$0.00 &      88 &     102 &      30 &     219 \\
Fe I  & 3631.48 & 0.96 & $+$0.00 &     102 &     118 &      23 &     168 \\
Fe I  & 3647.85 & 0.91 & $-$0.14 &      82 &      95 &      22 &     162 \\
Fe I  & 3679.92 & 0.00 & $-$1.58 &      69 &      80 &      17 &     125 \\
Fe I  & 3687.47 & 0.86 & $-$0.80 &      87 &     102 &      11 &      77 \\
Fe I  & 3727.63 & 0.96 & $-$0.61 &      83 &      96 &      10 &      75 \\
Fe I  & 3745.57 & 0.09 & $-$0.77 &     100 &     116 &      25 &     176 \\
Fe I  & 3745.91 & 0.12 & $-$1.34 &      89 &     104 &      10 &      72 \\
Fe I  & 3758.24 & 0.96 & $-$0.01 &     100 &     116 &      21 &     149 \\
Fe I  & 3763.80 & 0.99 & $-$0.22 &      86 &      99 &      15 &     106 \\
Fe I  & 3787.89 & 1.01 & $-$0.84 &      69 &      80 &      10 &      71 \\
Fe I  & 3820.44 & 0.86 & $+$0.16 &     114 &     133 &      22 &     157 \\
Fe I  & 3825.89 & 0.92 & $-$0.02 &      99 &     116 &      22 &     153 \\
Fe I  & 3840.45 & 0.99 & $-$0.50 &      74 &      87 &      20 &     140 \\
Fe I  & 3856.38 & 0.05 & $-$1.28 &      84 &      98 &      11 &      76 \\
Fe I  & 3865.53 & 1.01 & $-$0.95 &      67 &      79 &      11 &      76 \\
Fe I  & 3899.72 & 0.09 & $-$1.51 &      70 &      81 &      19 &     135 \\
Fe I  & 3917.18 & 0.99 & $-$2.15 &      27 &      31 &       4 &      27 \\
Fe I  & 3922.92 & 0.05 & $-$1.63 &      75 &      87 &      16 &     110 \\
Fe I  & 3949.96 & 2.18 & $-$1.25 &      18 &      21 &       6 &      38 \\
Fe I  & 4005.25 & 1.56 & $-$0.58 &      61 &      71 &      12 &      85 \\
Fe I  & 4071.75 & 1.61 & $-$0.01 &      79 &      92 &      17 &     125 \\
Fe I  & 4134.69 & 2.83 & $-$0.65 &      19 &      22 &       4 &      29 \\
Fe I  & 4143.88 & 1.56 & $-$0.51 &      68 &      80 &       8 &      53 \\
Fe I  & 4147.68 & 1.49 & $-$2.07 &      12 &      14 &       4 &      29 \\
Fe I  & 4175.64 & 2.85 & $-$0.83 &      18 &      21 & \nodata & \nodata \\
Fe I  & 4187.05 & 2.45 & $-$0.51 &      38 &      45 &       2 &      16 \\
Fe I  & 4187.81 & 2.42 & $-$0.51 &      42 &      49 &       6 &      41 \\
Fe I  & 4191.44 & 2.47 & $-$0.67 &      26 &      30 &       8 &      54 \\
Fe I  & 4199.10 & 3.05 & $+$0.16 &      52 &      61 & \nodata & \nodata \\
Fe I  & 4202.04 & 1.49 & $-$0.69 &      62 &      72 &      13 &      85 \\
Fe I  & 4216.19 & 0.00 & $-$3.36 &      24 &      28 &       2 &      13 \\
Fe I  & 4222.22 & 2.45 & $-$0.91 &      24 &      28 & \nodata & \nodata \\
Fe I  & 4233.61 & 2.48 & $-$0.58 &      37 &      43 &       4 &      24 \\
Fe I  & 4250.13 & 2.47 & $-$0.38 &      47 &      56 & \nodata & \nodata \\
Fe I  & 4260.48 & 2.40 & $+$0.08 &      61 &      72 &      10 &      69 \\
Fe I  & 4271.16 & 2.45 & $-$0.34 &      58 &      69 &      20 &     131 \\
Fe I  & 4271.77 & 1.49 & $-$0.17 &      93 &     109 &      11 &      77 \\
Fe I  & 4282.40 & 2.18 & $-$0.78 &      34 &      39 &       8 &      52 \\
Fe I  & 4383.55 & 1.48 & $+$0.21 &     104 &     122 &      15 &     100 \\
Fe I  & 4404.75 & 1.56 & $-$0.15 &      76 &      90 &      14 &      92 \\
Fe I  & 4415.12 & 1.61 & $-$0.62 &      62 &      73 &      16 &     108 \\
Fe I  & 4427.32 & 0.05 & $-$2.92 &      37 &      44 &      10 &      65 \\
Fe I  & 4430.61 & 2.22 & $-$1.73 &      10 &      12 &       2 &      15 \\
Fe I  & 4447.72 & 2.22 & $-$1.34 &      19 &      22 &       4 &      27 \\
Fe I  & 4461.65 & 0.09 & $-$3.19 &      20 &      24 &       8 &      52 \\
Fe I  & 4466.55 & 2.83 & $-$0.60 &      19 &      22 &       9 &      59 \\
Fe I  & 4489.73 & 0.12 & $-$3.90 &       6 &       6 &       3 &      20 \\
Fe I  & 4494.56 & 2.20 & $-$1.14 &      23 &      27 &       3 &      20 \\
Fe I  & 4528.61 & 2.18 & $-$0.89 &      35 &      42 &       8 &      51 \\
Fe I  & 4602.95 & 1.49 & $-$2.21 &      13 &      15 &       2 &      16 \\
Fe I  & 4871.32 & 2.87 & $-$0.36 &      25 &      29 &       5 &      34 \\
Fe I  & 4872.14 & 2.88 & $-$0.57 &      25 &      30 & \nodata & \nodata \\
Fe I  & 4890.76 & 2.88 & $-$0.39 &      24 &      29 &      10 &      65 \\
Fe I  & 4891.50 & 2.85 & $-$0.11 &      37 &      44 &       8 &      49 \\
Fe I  & 4903.32 & 2.88 & $-$0.93 &      12 &      14 & \nodata & \nodata \\
Fe I  & 4919.00 & 2.87 & $-$0.34 &      29 &      34 &       6 &      35 \\
Fe I  & 5006.12 & 2.83 & $-$0.61 &      23 &      28 & \nodata & \nodata \\
Fe I  & 5014.94 & 3.94 & $-$0.27 &      12 &      15 &       4 &      22 \\
Fe I  & 5049.83 & 2.28 & $-$1.35 &      18 &      21 &       8 &      49 \\
Fe I  & 5133.69 & 4.18 & $+$0.20 &      11 &      13 &       7 &      45 \\
Fe I  & 5191.47 & 3.04 & $-$0.55 &      21 &      25 & \nodata & \nodata \\
Fe I  & 5192.35 & 3.00 & $-$0.42 &      30 &      36 & \nodata & \nodata \\
Fe I  & 5194.95 & 1.56 & $-$2.02 &      17 &      25 & \nodata & \nodata \\
Fe I  & 5202.34 & 2.18 & $-$1.84 &      11 &      13 & \nodata & \nodata \\
Fe I  & 5216.28 & 1.61 & $-$2.08 &      16 &      19 & \nodata & \nodata \\
Fe I  & 5232.95 & 2.94 & $-$0.06 &      39 &      48 &       3 &      29 \\
Fe I  & 5266.56 & 3.00 & $-$0.39 &      24 &      29 &       4 &      21 \\
Fe I  & 5269.55 & 0.86 & $-$1.33 &      70 &      86 &      15 &      98 \\
Fe I  & 5283.63 & 3.24 & $-$0.52 &      17 &      20 &       3 &      20 \\
Fe I  & 5324.19 & 3.21 & $-$0.10 &      25 &      30 & \nodata & \nodata \\
Fe I  & 5328.05 & 0.92 & $-$1.47 &      68 &      81 &      13 &      82 \\
Fe I  & 5371.50 & 0.96 & $-$1.64 &      55 &      66 &      12 &      77 \\
Fe I  & 5383.38 & 4.31 & $+$0.65 &      17 &      20 & \nodata & \nodata \\
Fe I  & 5397.14 & 0.92 & $-$1.98 &      47 &      56 &       5 &      32 \\
Fe I  & 5405.78 & 0.99 & $-$1.85 &      44 &      52 &       9 &      61 \\
Fe I  & 5429.71 & 0.96 & $-$1.88 &      46 &      56 &      10 &      59 \\
Fe I  & 5434.53 & 1.01 & $-$2.13 &      36 &      44 &       7 &      40 \\
Fe I  & 5446.92 & 0.99 & $-$1.91 &      39 &      49 &      20 &     106 \\
Fe I  & 5501.48 & 0.96 & $-$3.05 &      10 &      12 & \nodata & \nodata \\
Fe I  & 5506.79 & 0.99 & $-$2.79 &      15 &      18 &       5 &      29 \\
Fe I  & 5569.63 & 3.42 & $-$0.50 &      11 &      14 &       3 &      17 \\
Fe I  & 5572.85 & 3.40 & $-$0.28 &      15 &      19 & \nodata & \nodata \\
Fe I  & 5586.77 & 3.37 & $-$0.10 &      21 &      25 &       5 &      30 \\
Fe I  & 5615.66 & 3.33 & $+$0.05 &      24 &      28 &       5 &      27 \\
Fe I  & 6136.62 & 2.45 & $-$1.41 &      16 &      19 & \nodata & \nodata \\
Fe I  & 6191.57 & 2.43 & $-$1.42 &      13 &      16 & \nodata & \nodata \\
Fe I  & 6230.74 & 2.56 & $-$1.28 &      16 &      19 &       3 &      17 \\
Fe I  & 6393.61 & 2.43 & $-$1.58 &      13 &      16 & \nodata & \nodata \\
Fe I  & 6400.01 & 3.60 & $-$0.29 &      13 &      16 & \nodata & \nodata \\
Fe I  & 6430.85 & 2.18 & $-$1.95 &      10 &      12 & \nodata & \nodata \\
Fe I  & 6494.98 & 2.40 & $-$1.24 &      22 &      26 &      12 &      66 \\
Fe II & 4178.86 & 2.57 & $-$2.48 &      20 &      23 & \nodata & \nodata \\
Fe II & 4233.17 & 2.58 & $-$2.00 &      37 &      44 &       3 &      19 \\
Fe II & 4508.28 & 2.84 & $-$2.22 &      16 &      19 & \nodata & \nodata \\
Fe II & 4555.89 & 2.82 & $-$2.28 &      12 &      14 &       5 &      33 \\
Fe II & 4923.92 & 2.89 & $-$1.32 &      48 &      56 &       8 &      47 \\
Fe II & 5234.62 & 3.22 & $-$2.05 &      17 &      20 & \nodata & \nodata \\
Fe II & 5275.99 & 3.20 & $-$1.94 &      15 &      18 & \nodata & \nodata \\
Co I  & 3502.29 & 0.43 & $+$0.07 &      52 &      61 &      11 &      81 \\
Co I  & 3506.33 & 0.51 & $-$0.04 &      39 &      46 &       9 &      65 \\
Ni I  & 3472.56 & 0.11 & $-$0.82 &      55 &      64 &      11 &      79 \\
Ni I  & 3483.78 & 0.27 & $-$1.12 &      50 &      57 & \nodata & \nodata \\
Ni I  & 3500.86 & 0.17 & $-$1.37 &      54 &      62 &      12 &      89 \\
Ni I  & 3510.33 & 0.21 & $-$0.65 &      73 &      84 &      12 &      90 \\
Ni I  & 3515.07 & 0.11 & $-$0.26 &      70 &      81 &      17 &     124 \\
Ni I  & 3519.76 & 0.27 & $-$1.42 &      36 &      41 &       5 &      37 \\
Ni I  & 3524.54 & 0.03 & $-$0.03 &      84 &      97 &      16 &     118 \\
Ni I  & 3566.39 & 0.42 & $-$0.27 &      64 &      74 &      12 &      84 \\
Ni I  & 3597.71 & 0.21 & $-$1.09 &      44 &      51 &      12 &      87 \\
Ni I  & 3619.40 & 0.42 & $-$0.04 &      82 &      95 &       7 &      50 \\
Sr II\tablenotemark{a} & 4077.71 & 0.00 & $+$0.17 &    (86) &   (101) &    (16) &   (110) \\
Sr II\tablenotemark{a} & 4215.52 & 0.00 & $-$0.17 &    (80) &    (94) &    (15) &    (99) \\
Y II  & 3611.04 & 0.13 & $+$0.01 & \nodata & \nodata & \nodata & \nodata \\
Y II  & 3710.29 & 0.18 & $+$0.46 & \nodata & \nodata & \nodata & \nodata \\
Y II  & 3774.33 & 0.13 & $+$0.21 & \nodata & \nodata & \nodata & \nodata \\
Zr II & 4208.99 & 0.71 & $-$0.46 & \nodata & \nodata & \nodata & \nodata \\
Ba II\tablenotemark{a} & 4554.03 & 0.00 & $+$0.17 &    (92) &   (108) &    (13) &    (82) \\
Ba II\tablenotemark{a} & 4934.10 & 0.00 & $-$0.16 &    (82) &    (98) &    (10) &    (62) \\
Ba II & 5853.69 & 0.60 & $-$0.91 & \nodata & \nodata & \nodata & \nodata \\
Ba II\tablenotemark{a} & 6141.73 & 0.70 & $-$0.08 &    (58) &    (71) &     (8) &    (46) \\
Ba II\tablenotemark{a} & 6496.91 & 0.60 & $-$0.38 &    (53) &    (65) &     (7) &    (36) \\
La II & 3988.51 & 0.17 & $+$0.21 & \nodata & \nodata & \nodata & \nodata \\
La II & 3995.74 & 0.17 & $-$0.06 & \nodata & \nodata & \nodata & \nodata \\
La II & 4086.71 & 0.00 & $-$0.07 & \nodata & \nodata & \nodata & \nodata \\
La II & 4123.22 & 0.32 & $+$0.13 & \nodata & \nodata & \nodata & \nodata \\
Ce II & 4137.65 & 0.52 & $+$0.44 & \nodata & \nodata & \nodata & \nodata \\
Nd II & 4021.33 & 0.32 & $-$0.10 & \nodata & \nodata & \nodata & \nodata \\
Nd II & 4061.08 & 0.47 & $+$0.55 & \nodata & \nodata & \nodata & \nodata \\
Nd II & 4462.98 & 0.56 & $+$0.04 & \nodata & \nodata & \nodata & \nodata \\
Eu II & 3819.67 & 0.00 & $+$0.51 & \nodata & \nodata & \nodata & \nodata \\
Eu II & 3907.11 & 0.21 & $+$0.17 & \nodata & \nodata & \nodata & \nodata \\
Eu II & 3971.97 & 0.21 & $+$0.27 & \nodata & \nodata & \nodata & \nodata \\
Eu II & 4129.72 & 0.00 & $+$0.22 & \nodata & \nodata & \nodata & \nodata \\
Eu II & 4205.05 & 0.00 & $+$0.21 & \nodata & \nodata & \nodata & \nodata \\
Yb II & 3694.17 & 0.00 & $-$0.30 & \nodata & \nodata & \nodata & \nodata \\
Pb I  & 4057.81 & 1.32 & $-$0.17 & \nodata & \nodata & \nodata & \nodata \\

\enddata

\tablenotetext{a}{$EW$ values are given for illustration only, and were
                  not used in final abundance determinations.}

\end{deluxetable}

\clearpage

\begin{deluxetable}{lrrrrrrrrrrrrr}
%\rotate
\tablecolumns{13}
\tablewidth{0pt}
\tabletypesize{\footnotesize}
\tablecaption{Individual Abundance Results for Primary and Secondary 
              Stars\label{tab4}}
\tablehead{
\colhead{Species}                   &
\colhead{log $\epsilon_{\odot}$}    &
\colhead{log $\epsilon_p$}          &
\colhead{$\sigma_p$}                &
\colhead{\#$_p$}                    &
\colhead{log $\epsilon_s$}          &
\colhead{$\sigma_s$}                &
\colhead{\#$_s$}                    &
\colhead{[X/H]$_p$}                 &
\colhead{[X/H]$_s$}                 &
\colhead{[X/Fe]$_p$}                &
\colhead{[X/Fe]$_s$}                &
\colhead{Method}                    &
}
 
\startdata
CH    & 8.70 & 7.64 &    0.10 & \nodata & 7.44 &    0.40 & \nodata & $-$1.06 & $-$0.86 & $+$1.35 & $+$1.15 & syn \\
Fe I  & 7.52 & 5.13 &    0.16 &     104 & 5.11 &    0.44 &      83 & $-$2.39 & $-$2.41 & $+$0.02 &    0.00 &  EW \\
Fe II & 7.52 & 5.09 &    0.12 &       7 & 5.11 &    0.44 &       3 & $-$2.43 & $-$2.41 & $-$0.02 &    0.00 &  EW \\
Na I  & 6.33 &\nodata&\nodata&  \nodata & 3.9  &    0.3  &       2 &\nodata  & $-$2.4  &  \nodata&    0.0  & syn \\
Mg I  & 7.58 & 5.53 &    0.20 &       6 & 5.52 &    0.15 &       6 & $-$2.05 & $-$2.06 & $+$0.36 & $+$0.35 &  EW \\
Al I  & 6.57 & 3.26 & \nodata &       1 & 3.28 & \nodata &       1 & $-$3.31 & $-$3.29 & $-$0.90 & $-$0.88 &  EW \\
Ca I  & 6.36 & 4.38 &    0.13 &       8 & 4.03 &    0.29 &       6 & $-$1.98 & $-$2.33 & $+$0.43 & $+$0.08 &  EW \\
Ti II & 4.99 & 3.19 &    0.26 &       9 & 3.14 &    0.75 &       4 & $-$1.80 & $-$1.85 & $+$0.61 & $+$0.56 &  EW \\
Cr I  & 5.64 & 3.36 &    0.30 &       8 & 2.81 &    0.53 &       6 & $-$2.28 & $-$2.83 & $+$0.13 & $-$0.42 &  EW \\
Cr II & 5.64 & 3.45 &    0.10 &       2 & 3.64 & \nodata &       1 & $-$2.19 & $-$2.00 & $+$0.22 & $+$0.41 &  EW \\
Ni I  & 6.25 & 3.92 &    0.24 &      10 & 3.86 &    0.62 &       9 & $-$2.33 & $-$2.39 & $+$0.08 & $+$0.02 &  EW \\
Sr II & 2.90 & 1.05 &    0.09 &       2 & 0.93 &    0.29 &       2 & $-$1.85 & $-$1.97 & $+$0.56 & $+$0.44 & syn \\
Ba II & 2.13 & 1.17 &    0.08 &       1 & 1.09 &    0.28 &       1 & $-$0.96 & $-$1.04 & $+$1.45 & $+$1.37 & syn \\
\enddata
 
\end{deluxetable}

\clearpage

\begin{deluxetable}{lrrrrrrrrrrrrr}
%\rotate
\tablecolumns{13}
\tablewidth{0pt}
\tabletypesize{\normalsize}
\tablecaption{Abundance for the CS~22964-161 System\label{tab5}}
\tablehead{
\colhead{Species}                        &
\colhead{log $\epsilon_{\odot}$}         &
\colhead{log $\epsilon$}                 &
\colhead{$\sigma$}                       &
\colhead{\#}                             &
\colhead{[X/H]}                          &
\colhead{[X/Fe]}                         &
\colhead{Method\tablenotemark{(a)}}      &
}

\startdata
Fe I  &    7.52 & $+$5.13 & 0.21 & 104 & $-$2.39 & $+$0.02  & est \\
Fe II &    7.52 & $+$5.09 & 0.17 &   7 & $-$2.43 & $-$0.02  & est \\
Li I  & \nodata & $+$2.09 & 0.20 &   1 & \nodata & \nodata  & est \\
CH    &    8.70 & $+$7.50 & 0.15 &   1 & $-$1.20 & $+$1.21  & syn \\
Na I  &    6.33 & $+$3.9  & 0.3  &   4 & $-$2.4  &    0.0   & syn \\
Mg I  &    7.58 & $+$5.53 & 0.20 &   6 & $-$2.05 & $+$0.36  & syn \\
Al I  &    6.57 & $+$3.26 & 0.20 &   1 & $-$3.31 & $-$0.90  & est \\
Si I  &    7.55 & $+$5.28 & 0.25 &   1 & $-$2.27 & $+$0.14  & syn \\
Ca I  &    6.36 & $+$4.32 & 0.16 &   8 & $-$2.04 & $+$0.37  & est \\
Ti II &    4.99 & $+$3.18 & 0.34 &   9 & $-$1.81 & $+$0.60  & est \\
Cr I  &    5.64 & $+$3.27 & 0.34 &   8 & $-$2.37 & $+$0.04  & est \\
Cr II &    5.64 & $+$3.48 & 0.12 &   2 & $-$2.16 & $+$0.25  & est \\
Mn I  &    5.39 & $+$2.27 & 0.15 &   3 & $-$3.12 & $-$0.71  & syn \\
Ni I  &    6.25 & $+$3.91 & 0.30 &  10 & $-$2.34 & $+$0.07  & est \\
Sr II &    2.90 & $+$1.01 & 0.11 &   2 & $-$1.89 & $+$0.52  & syn \\
Y II  &    2.24 & $+$0.12 & 0.10 &   3 & $-$2.12 & $+$0.29  & syn \\
Zr II &    2.60 & $+$0.88 & 0.25 &   1 & $-$1.72 & $+$0.69  & syn \\
Ba II &    2.13 & $+$0.98 & 0.16 &   5 & $-$1.15 & $+$1.26  & syn \\
La II &    1.13 & $-$0.21 & 0.10 &   4 & $-$1.34 & $+$1.07  & syn \\
Ce II &    1.55 & $+$0.43 & 0.25 &   1 & $-$1.12 & $+$1.29  & syn \\
Nd II &    1.45 & $+$0.05 & 0.12 &   3 & $-$1.40 & $+$1.01  & syn \\
Eu II &    0.52 & $-$1.20 & 0.26 &   5 & $-$1.72 & $+$0.69  & syn \\
Yb II &    1.08 & $-$0.69 & 0.25 &   1 & $-$1.77 & $+$0.64  & syn \\
Pb I  &    1.85 & $+$1.63 & 0.20 &   1 & $-$0.22 & $+$2.19  & syn \\
\enddata

\tablenotetext{(a)}{Abundance method: ``est'' means that the system abundance
was estimated from the individual primary and secondary abundances given
in Table~\ref{tab4}; ``syn'' means that the system abundance was
computed with synthetic spectra of features in the syzygy spectrum.}

\end{deluxetable}

\clearpage

\begin{deluxetable}{@{\extracolsep{0.15in}}lcccc}
\tablecolumns{5}
\tablewidth{0pt}
\tabletypesize{\normalsize}
\tablecaption{Carbon-Rich Star Periods and Velocity Ranges\label{tab6}}
\tablehead{
\colhead{Group}                        &
\colhead{P\tablenotemark{(a)} range}   &
\colhead{Median P}                     &
\colhead{K\tablenotemark{(b)} range}   &
\colhead{Median K}                     \\
\colhead{}                             &
\colhead{days}                         &
\colhead{days}                         &
\colhead{km s$^{-1}$}                  &
\colhead{km s$^{-1}$}                  
}

\startdata
Barium-CH giant &   80 $-$ 4390 & 1352 &  2.5 $-$ 12.0 & 6.0 \\
CH subgiant     &  880 $-$ 4140 & 1930 &  3.3 $-$  7.2 & 4.6 \\
\enddata

\tablenotetext{(a)}{orbital period}
\tablenotetext{(b)}{velocity semi-amplitude}

\end{deluxetable}

\clearpage

\begin{deluxetable}{lll}
\tablecolumns{3}
\tablewidth{0pt}
\tabletypesize{\normalsize}
\tablecaption{Predicted mass fraction and molecular weight\label{tab7}}
\tablehead{
\colhead{Mass Fraction}                 &
\colhead{$M$=1.3\Msun}                  &
\colhead{$M$=1.5\Msun}                     
}
                                                                                
\startdata
\cutinhead{At last Third Dredge-Up}
$X$     & 0.675  & 0.535\\
$Y$     & 0.301  & 0.403\\
$Z$     & 0.0238 & 0.062\\
$\mu$   & 0.630  & 0.713\\
\cutinhead{Mass average of the winds}
$X$     & 0.700  & 0.585\\
$Y$     & 0.284  & 0.367\\
$Z$     & 0.0155 & 0.0483\\
$\mu$   & 0.616  & 0.781\\
\enddata
                                                                                
\end{deluxetable}

\end{document}